\documentclass[journal]{IEEEtran}

\usepackage{cite}
\usepackage{color}
\usepackage{amsfonts,amssymb,amsmath,verbatim}
\usepackage{subfigure}
\usepackage{float}
\usepackage{graphicx}
\usepackage{boxedminipage}

\newcommand{\ff}{\mathbb{F}}
\newcommand{\cc}{\mathsf{C}}

\begin{document}

\title{CONDENSE: A Reconfigurable Knowledge Acquisition Architecture for Future 5G IoT}

\author{Dejan~Vukobratovic, Dusan~Jakovetic, Vitaly~Skachek, Dragana~Bajovic, Dino~Sejdinovic, Gunes~Karabulut~Kurt, Camilla~Hollanti, and~Ingo~Fischer
\thanks{D.~Vukobratovic is with the Department of Power, Electronics and Communications Engineering, University of Novi Sad, Serbia, e-mail: dejanv@uns.ac.rs.}
\thanks{D.~Jakovetic is with the BioSense Institute, Novi Sad, Serbia, and with the Department of Mathematics and Informatics, Faculty of Sciences, University of Novi Sad, Serbia, email: djakovet@uns.ac.rs.}
\thanks{V.~Skachek is with the Institute of Computer Science, University of Tartu, Estonia, email: vitaly.skachek@ut.ee.}%
\thanks{D. Bajovic is with the BioSense Institute, Novi Sad, Serbia, and with the Department of Power, Electronics and Communication Engineering, University of Novi Sad, Serbia, email: dbajovic@uns.ac.rs.}
\thanks{D.~Sejdinovic is with the Department of Statistics, University of Oxford, UK, email: dino.sejdinovic@stats.ox.ac.uk.}
\thanks{G.~Karabulut Kurt is with the Department of Electronics and Communication Engineering, Istanbul Technical University, Turkey, email: gkurt@itu.edu.tr.}
\thanks{C.~Hollanti is with the Department of Mathematics and Systems Analysis, Aalto University, Finland, email: camilla.hollanti@aalto.fi.}
\thanks{I.~Fischer is with the Institute for Cross-Disciplinary Physics and Complex Systems (UIB-CSIC), Spain, email: ingo@ifisc.uib-csic.es.}
\thanks{D. Vukobratovic is financially supported by Rep. of Serbia TR III 44003 grant. V. Skachek is supported in part by the grant PUT405 from the Estonian Research Council. G. Karabulut Kurt is supported by TUBITAK Grant 113E294. C. Hollanti is financially supported by the Academy of Finland grants \#276031, \#282938 and \#283262.}}

\maketitle

\begin{abstract}
In forthcoming years, the Internet of Things (IoT) will connect billions of smart devices  generating and uploading a deluge of data to the cloud. If successfully extracted, the knowledge buried in the data can significantly improve the quality of life and foster economic growth. However, a critical bottleneck for realising the efficient IoT is the pressure it puts on the existing communication infrastructures, requiring transfer of enormous data volumes. Aiming at addressing this problem, we propose a novel architecture dubbed Condense (re\underline{con}figurable knowle\underline{d}g\underline{e} acquisitio\underline{n} \underline{s}yst\underline{e}ms), which integrates the IoT-communication infrastructure into data analysis. This is achieved via the generic concept of network function computation: Instead of merely transferring data from the IoT sources to the cloud, the communication infrastructure should actively participate in the data analysis by carefully designed en-route processing. We define the Condense architecture, its basic layers, and the interactions among its constituent modules. Further, from the implementation side, we describe how Condense can be integrated into the 3rd Generation Partnership Project (3GPP) Machine Type Communications (MTC) architecture, as well as the prospects of making it a practically viable technology in a short time frame, relying on Network Function Virtualization~(NFV) and Software Defined Networking~(SDN). Finally, from the theoretical side, we survey the relevant literature on computing ``atomic'' functions in both analog and digital domains, as well as on function decomposition over networks, highlighting challenges, insights, and future directions for exploiting these techniques within practical 3GPP MTC architecture.
\end{abstract}


\IEEEpeerreviewmaketitle

\section{Introduction}

A deluge of data is being generated by an ever-increasing number of devices that indiscriminately collect, process and upload data to the cloud. An estimated 20 to 40 billion devices will be connected to the Internet by 2020 as part of the Internet of Things (IoT)~\cite{Siemens}. IoT has the ambition to interconnect smart devices across cities, vehicles, appliances, connecting industries, retail and healthcare domains, thus becoming a dominant fuel for the emerging Big Data revolution \cite{Time}. IoT is considered as one of the key technologies to globally improve the quality of life, economic growth, and employment, with the European Union market value expected to exceed one trillion Euros in 2020~\cite{EU}. However, a critical bottleneck for the IoT vision is the pressure it puts on the existing communication infrastructures, by requiring transfer of enormous amounts of data. By 2020 IoT data will exceed 4.4~ZB (zettabytes) amounting to 10$\%$ of the global ``digital universe'' (compared to 2$\%$ in 2013)~\cite{ZDNet}. Therefore, a sustainable solution for IoT and cloud integration is one of the main challenges for contemporary communications technologies.

The state-of-the-art in IoT/cloud integration assumes uploading and storing all the raw data generated by IoT devices to the cloud. The IoT data is subsequently processed by cloud-based data analysis that aims to extract useful knowledge~\cite{IoTCloud}. For majority of applications, this approach is inefficient since there is typically a large amount of redundancy in the collected data. As a preprocessing step prior to data analysis, projections to a much lower-dimensional space are often employed, essentially discarding large portions of data. With the growth of IoT traffic, the approach where communications and data analysis are separated will become unsustainable, necessitating a  fundamental redesign of IoT communications.

In this work, we propose a generic and reconfigurable IoT architecture capable of adapting the IoT data transfer to the subsequent data analysis. We refer to the proposed architecture as Condense (re\underline{con}figurable knowle\underline{d}g\underline{e} acquisitio\underline{n} \underline{s}yst\underline{e}ms). Instead of merely transferring data, the proposed architecture provides an active and reconfigurable service leveraged by the data analysis process. We identify a common generic interface between data communication and data analysis: \emph{the function computation}, and we distinguish it as a core Condense technology. Instead of communicating a stream of data units from the IoT devices to the cloud, the proposed IoT architecture processes the data units \emph{en-route} through a carefully designed process to deliver a stream of network function evaluations stored in the cloud. In other words, the Condense architecture does not transfer all the raw data across the communications infrastructure, but only what is needed from the perspective of the current application at hand.

To illustrate the idea with a toy example, consider a number of sensors which constitute a fire alarm system, e.g.,~\cite{Giridhar05},~\cite{Giridhar}. Therein, we might only be interested in the maximal temperature across the sensed field, and not in the full sensors readings vector. Therefore, it suffices to deliver to the relevant cloud application only an evaluation of the maximum function applied over the sensors readings vector; Condense realizes this maximum function as a composition of ``atomic'' functions implemented across the communications infrastructure.

We describe how to implement the proposed approach explained above in the concrete third generation partnership project (3GPP) Machine Type Communications (MTC) architecture~\cite{Taleb2012}. The 3GPP MTC service is expected to contribute a dominant share of the IoT traffic via the upcoming fifth generation~(5G) mobile cellular systems, thus providing an ideal setup for the demonstration of Condense concepts. We enhance the 3GPP MTC architecture with the network function computation~(NFC) -- a novel envisioned MTC-NFC service. We define the layered Condense architecture comprised of three layers: i) atomic function computation layer, ii) network function computation layer, and iii) application layer, and we map these layers onto the 3GPP MTC architecture. In the lowermost atomic function computation (AFC) layer, carefully selected atomic modules perform local function computations over the input data. The network function computation layer orchestrates the collection of AFC modules into the global network-wide NFC functionality, thus evaluating non-trivial functions of the input data as a coordinated composition of AFCs. Furthermore, the NFC layer provides a flexible and reconfigurable MTC-NFC service to the topmost application layer, where cloud-based data analysis applications directly exploit the outputs of the NFC layer. Throughout the system description, we provide a review of the theoretical foundations that justify the proposed architecture and point to the tools for the system design and analysis. Finally, we detail practical viability of incorporating NFC services within 3GPP MTC service, relying on emerging concepts of Network Function Virtualization (NFV)~\cite{etsiNFV},~\cite{Han2015} and Software Defined Networking (SDN)~\cite{Lantz2010},~\cite{Kreutz2015}; this upgrade is, thanks to the current uptake of the SDN/NFV concepts, achievable within a short time frame. 

This paper is somewhat complementary with respect to other works that consider architectures for 5G IoT communications. For example, reference \cite{Condoluci} focuses on machine-type multicast services to ensure end-to-end reliability, low latency and low energy consumption of MTC traffic (including both up and downlinks). Reference \cite{Palattella} provides a detailed analysis of integration of 5G technologies for the future global IoT, both from technological and standardization aspects. However, while existing works consider making communication of the MTC-generated \emph{raw data} efficient, here we aim to improve the overall system efficiency through communicating over the network infrastructure only the application-requested \emph{functions over data}.
In other words, this paper describes how we can potentially exploit decades of research on function computation and function decomposition over networks within the concrete, practical and realizable knowledge acquisition system for the IoT-generated data. In particular, we review the main results on realizing (atomic) function computation in the analog (wireless and optical) and digital domains, as well as on function evaluation and decomposition over networks, including the work on sensor fusion, e.g.,~\cite{Giridhar05,Giridhar}, network coding for computing~\cite{AFKZ,Kowshik2012}{-}\hspace{-1.6mm}~\cite{Dougherty}, and neural networks~\cite{Li2014}{-}{\hspace{-1.6mm}}~\cite{Srivastava2014}. While this paper does not provide novel contributions to these fields, it identifies and discusses main challenges in applying them within the practical 3GPP MTC architecture, and it points to interesting future research directions.

\textbf{Paper organization}. The rest of the paper is organized as follows. In Sec.~{II}, we review the state-of-the-art 3GPP MTC architecture, briefly present SDN/NFV concepts, and give notational conventions. In Sec.~{III}, we introduce the novel layered Condense architecture that, through the rest of the paper, we integrate into the 3GPP MTC architecture. In Sec.~{IV}, we describe the atomic function computation layer that defines the basic building block of the architecture, distinguishing between the analog (or \emph{in-channel}) AFC and digital (\emph{in-node}) AFC modules. The theoretical fundamentals and practical aspects of the NFC layer are presented in Sec.~{V}. In Sec.~{VI}, the interaction between the application layer and the NFC layer is discussed, where several application layer examples are presented in detail. Further implementation issues are discussed in Sec.~{VII}, and the paper is concluded in Sec.~{VIII}.

\section{Background and Preliminaries}

Subsection~{II-A} reviews the current 3GPP MTC architecture, Subsection~{II-B} gives background on software defined networking~(SDN) and network function virtualization~(NFV), while Subsection~{II-C} defines notation used throughout the rest of the paper.

\subsection{The 3GPP MTC Architecture}

Machine Type Communications~(MTC) is an European Telecommunications Standards Institute~(ETSI)-defined architecture that enables participating devices to send data to each other or to a set of servers~\cite{Taleb2012}. While ETSI is responsible for defining the generic MTC architecture, specific issues related with mobile cellular networks are addressed in 3GPP standardization \cite{3gppmtc}. 3GPP MTC is first included in Release~{10} and will evolve beyond current 3GPP Long Term Evolution~(LTE)/LTE-Advanced Releases into the 5G system~\cite{mtc5g}.

\begin{figure}
\centering
\includegraphics[trim = 0mm 53mm 0mm 0mm, clip, width=3in]{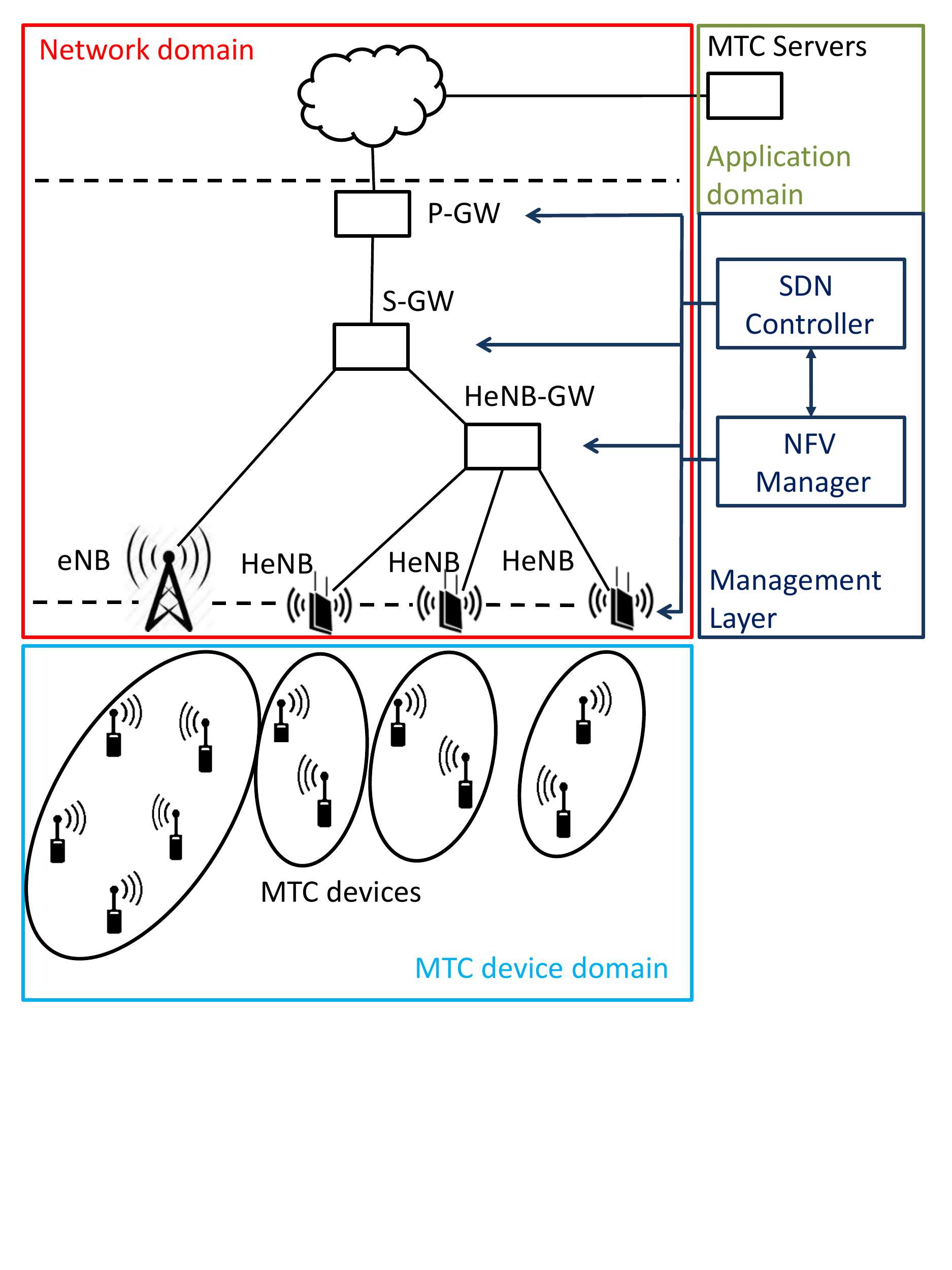}
\caption{The 3GPP MTC architecture.}
\label{3GPP-MTC}
\end{figure}

Fig.~\ref{3GPP-MTC} illustrates the 3GPP MTC architecture. It consists of: i) the MTC device domain containing MTC devices that access MTC service to send and/or receive data, ii) the network domain containing network elements that transfer the MTC device data, and iii) the MTC application domain containing MTC applications running on MTC servers. MTC devices access the network via Radio Access Network (RAN) elements: base stations (eNB: eNodeB) and small cells (HeNB: Home-eNodeB). Packet data flows follow the Evolved Packet Core (EPC) elements: HeNB Gateway (HeNB-GW), Service Gateway (S-GW) and Packet Gateway (P-GW), until they reach either a mobile operator MTC server or a third party MTC server via the Internet. In this work, we address MTC device data processing and focus on the data plane while ignoring the control plane of the 3GPP architecture.

Abstracted to its essence, the current 3GPP MTC approach in the context of IoT/cloud integration is represented by three layers (Fig.~1). The MTC device domain, or data layer, contains billions of devices that generate data while interacting with the environment. The network domain, or communication layer, provides mere data transfer services to the data layer by essentially uploading the generated data to the cloud in its entirety. The application domain, or application layer, contains data centres running MTC servers which provide storage and processing capabilities. MTC applications running in data centres enable, e.g., machine learning algorithms to extract knowledge from the collected data. In this paper, we challenge this 3GPP MTC layered structure and propose a novel Condense layered architecture described in Sec.~{III}.

\subsection{Software Defined Networking (SDN) and Network Function Virtualization (NFV)}

SDN and NFV are novel concepts in networking research that increase network flexibility and enable fast implementation of new services and architectures. Both SDN and NFV are under a current consideration for future integration in the 3GPP cellular architecture \cite{sdncell},~\cite{3gppnfv}. Although not yet part of the 3GPP architecture, in Fig. \ref{3GPP-MTC}, we present main NFV/SDN management entities: the NFV manager and the SDN controller, as they will be useful for the description of the Condense architecture.

SDN is a novel network architecture that decouples the control plane from the data plane \cite{Lantz2010},~\cite{sdnieee}. This is achieved by centralizing the traffic flow control where a central entity, called the SDN controller, manages the physical data forwarding process in SDN-enabled network nodes. The SDN controller remotely manages network nodes and flexibly controls traffic flows at various flow granularities. In other words, the SDN controller can easily (re)define forwarding rules for data flows passing through an SDN network node. Using SDN, various network services are able to quickly re-route their data flows, adapting the resulting (virtual) network topology to their needs.

NFV is another recent trend in networking where, instead of running various network functions (e.g., firewalls, NAT servers, load balancers, etc.) on dedicated network nodes, the network hardware is virtualized to support software-based implementations of network functions \cite{Han2015}. This makes network functions easy to instantiate anywhere across the network when needed. Multiple instances of network functions are jointly administered and orchestrated by the centralized NFV management.

NFV and SDN are complementary concepts that jointly provide flexible and efficient service chaining: a sequence of data processing tasks performed at different network nodes \cite{Li2015}. The NFV manager has the capability to actually instantiate the targeted (atomic) function computations at each node in the network. Similarly, SDN has the power to steer data flows and hence establish a desired (virtual) network topology which supports the desired network-wide computation. This feature will be fundamental for a fast implementation and deployment of the Condense architecture, as detailed in the rest of the paper. For more details about SDN/NFV concepts in 3GPP networks, we refer the interested reader to~\cite{sdncell}.

\subsection{Notational preliminaries}

Throughout, we use bold symbols to denote vectors, where $\ell$-th entry of a vector~$\mathbf{x}$ of length~$L$ is denoted by $x[\ell]$, $\ell=1,...,L$. We denote by $\mathbb R$ the set of real numbers, and by ${\mathbb R}^Q$ the $Q$-dimensional real coordinate space. A finite field is denoted by $\mathbb F$, a finite alphabet (finite discrete set) by $\mathbb A$, and by ${\mathbb A}^Q$ the set of $Q$-dimensional vectors with the entries from $\mathbb A$. Symbol $|\cdot|$ denotes the cardinality of a set. We deal with vectors $\mathbf x \in {\mathbb R}^Q$, $\mathbf x \in {\mathbb A}^Q$, and also $\mathbf x \in {\mathbb F}^Q$, and it is clear from context which of the three cases is in force. Also, addition and multiplication over both $\mathbb R$ and $\mathbb F$ are denoted in a standard way -- respectively as $+$ and $\cdot$ (or the multiplication symbol is simply omitted), and again the context clarifies which operation is actually applied.

We frequently consider a directed acyclic graph $\mathcal{G} = \left( \mathcal V, \mathcal E\right)$, where $\mathcal V$ denotes the set of nodes, and $\mathcal E$ the set of directed edges (arcs). An arc from node $u$ to node $v$ is denoted by
$u \rightarrow v$. Set $\mathcal V = \mathcal S \cup \mathcal A \cup \mathcal D$, where $\mathcal S$, $\mathcal A$, and $\mathcal D$ denote, respectively, the set of source nodes, atomic nodes, and destination nodes. We let $\mathcal S \cap \mathcal D = \varnothing$. We also introduce $N = |\mathcal S|$, $M = |\mathcal A|$, and $R = |\mathcal D|$. As we will see further ahead, source nodes correspond to MTC devices ($N$ data generators), atomic nodes correspond to the 3GPP communication infrastructure nodes which implement atomic functions ($M$ atomic nodes), and destination nodes are MTC servers in data centers which are to receive the desired function computation results ($R$ destination nodes). We index an arbitrary node in $\mathcal S$ by $s$, and similarly we write $a \in \mathcal A$, and $d \in \mathcal D$. When we do not intend to make a distinction among $\mathcal S, \mathcal A$, and $\mathcal D$, we index an arbitrary node by $v \in \mathcal V$. For each node $v \in \mathcal V$, we denote by $\mathcal V^{(v)}_{\mathrm{in}}$ its in-neighborhood, i.e., the set of nodes~$v^\prime$ in $\mathcal V$ such that the arc $v^\prime \rightarrow v$ exists. Analogously, $\mathcal V^{(v)}_{\mathrm{out}}$ denotes the node $v$'s out-neighborhood. As we frequently deal with in-neighborhoods, we will simply write $\mathcal V^{(v)}  \equiv \mathcal V^{(v)}_{\mathrm{in}}$. We call the in-degree of $v$ the cardinality of $\mathcal V^{(v)}_{\mathrm{in}}$, and we analogously define the out-degree. Although not required by the theory considered ahead, just for simplicity of notation and presentation, all sections except Section~V consider the special case where $\mathcal G$ is a directed rooted tree with $N$ sources and a single destination. Pictorially, we visualize $\mathcal G$ as having the source nodes $\mathcal S$ at the bottom, and the destination node $d$ at the top (see Sec. V, Fig. \ref{NFCgraph}, right-hand side). In the case of a directed rooted tree graph~$\mathcal G$, the leaf nodes' set of~$v$ coincides with its in-neighborhood $\mathcal{V}^{(v)}$, and all nodes except the destination nodes have the out-degree one, the destination node having the out-degree zero.

We index (vector) quantities associated with sources $s \in \mathcal S$ through subscripts, i.e., $\mathbf{x}_s$ is the source $s$'s vector. When considering a generic directed acyclic graph~$\mathcal G$ (Section~V), we associate to each
arc $u \rightarrow v$ a vector quantity $\mathbf{x}^{(u \rightarrow v)}$. With directed rooted trees (Sections III, IV, and VI), each node (except the destination node) has the out-degree one; hence, for simplicity, we then use node-wise (as opposed to edge-wise) notation, i.e., we index quantity $\mathbf{x}^{(u \rightarrow v)}$ as $\mathbf{x}^{(u)}$. Note that this notation is sufficient as, with directed rooted trees, there is only a single arc outgoing a (non-destination) node. When needed, time instances are denoted by $t=1,2,...,T$; a vector  associated with source $s$ and time $t$ is denoted by $\mathbf{x}_{s,t}$; similarly, we use $\mathbf{x}^{(v)}_t$ for non-source nodes.

\section{CONDENSE Architecture: IoT/Cloud Integration for 5G}

In this section, we present the Condense architecture that upgrades the 3GPP MTC architecture with the concept of network function computation (NFC). NFC creates a novel role 3GPP MTC service should offer: instead of communicating raw data, it should deliver function computations over the data, providing for a novel MTC-NFC service. The NFC design should be generic, flexible and reconfigurable to meet the needs of increasing number of MTC applications that extract knowledge from MTC data. For most applications, indiscriminate collection of MTC data is extremely wasteful and MTC-NFC service may dramatically reduce MTC traffic while preserving operational efficiency of MTC applications.

The Condense architecture challenges the conventional division into data, communications and application layer (Sec. 2A). Instead, we propose a novel architecture consisting of: i) atomic function computation (AFC) layer, ii) network function computation (NFC) layer, and iii) application layer. In this section, we provide a high-level modular description of the architecture by carefully defining its basic building blocks (modules). In the following three sections, we delve into details of each layer and provide both theoretical justifications and implementation discussion that motivated this work.

\subsection{CONDENSE Architecture: Modules and Layers}

\begin{figure}
\centering
\includegraphics[width=3.4in]{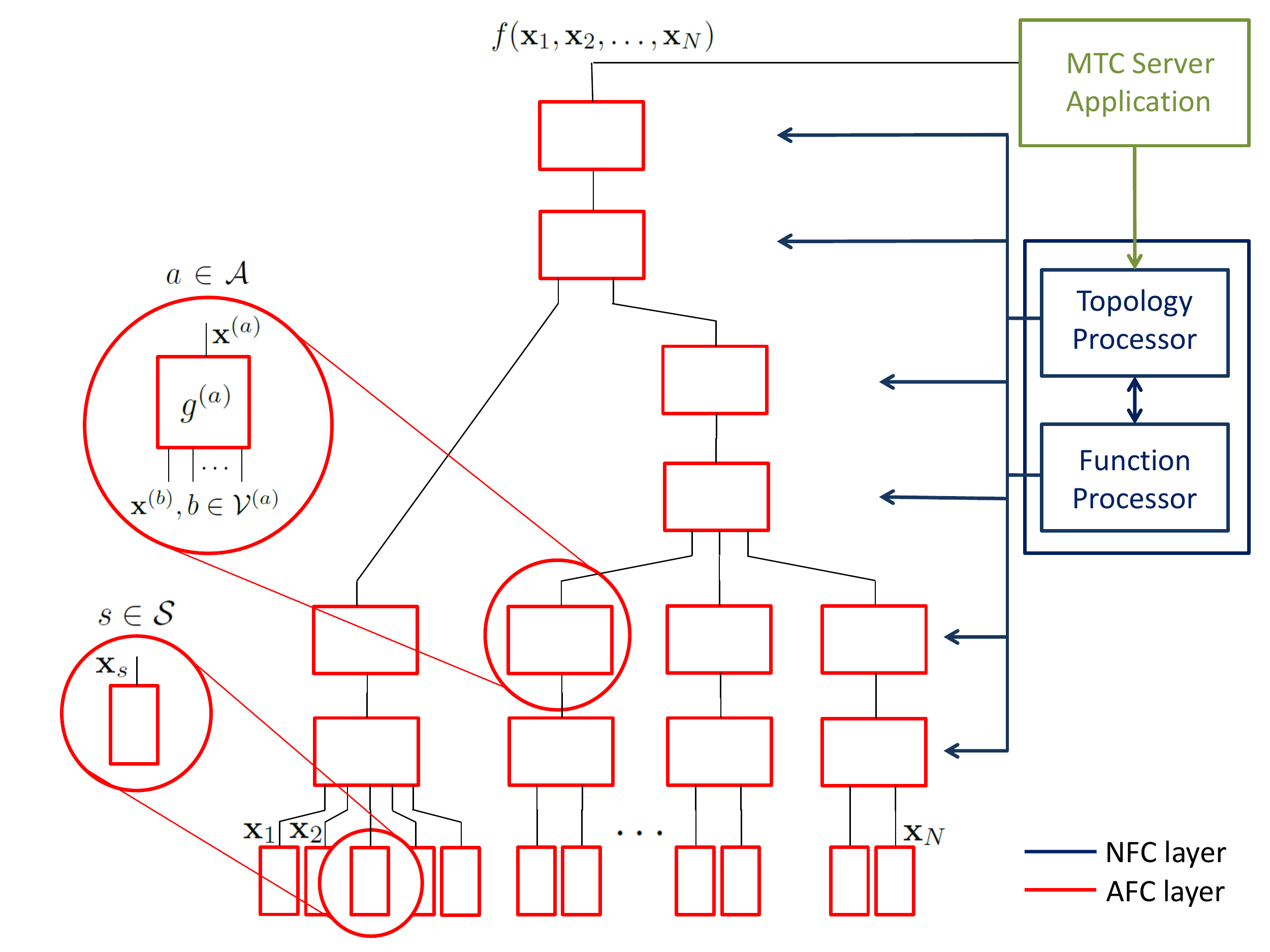}
\caption{Condense MTC-NFC architecture.}
\label{MTC-NFC}
\end{figure}

The Condense architecture is presented in Fig.~\ref{MTC-NFC}. It consists of an interconnected collection of basic building blocks called AFC modules. Each AFC module evaluates an (atomic) function over the input data packets and delivers an output data packet representing the atomic function evaluation. A generic AFC module may have multiple input and multiple output interfaces, each output interface representing a different AFC over the input data. The collection of interconnected and jointly orchestrated AFC modules delivers a network function computation over the source data packets. The resulting NFC evaluations are the input to application layer MTC server application.

Let us assume that an MTC network contains $N$ MTC devices representing the set of source modules (or source nodes) $\mathcal{S}$. Source node $s \in \mathcal{S}$ produces a message $\mathbf{x}_{s}=(x_{s}[1],x_{s}[2],\ldots,x_{s}[L])$ containing $L$ symbols from a given alphabet $\mathbb A$. The message $\mathbf{x}_{s}$ is transmitted at an output interface of the source module $s$. For simplicity, we assume that every source module has a single output interface.

In addition to the source nodes, the MTC network contains $M$ AFC modules (or AFC nodes) representing the set $\mathcal{A}$. An arbitrary AFC node $a \in \mathcal{A}$ has $P$ input and $Q$ output interfaces. For simplicity, unless otherwise stated, we will assume single-output AFC modules, i.e., $Q=1$. At input interfaces, the AFC node $a$ receives the set of input data packets $\{\mathbf{x}^{(b)}\}_{b \in \mathcal V^{(a)}}$, while at the output interface, it delivers the output data packet $\mathbf{x}^{(a)}$. AFC node~$a$ associates an atomic function $g^{(a)}$ to the output interface, where $\mathbf{x}^{(a)}=g^{(a)}(\{\mathbf{x}^{(b)}\}_{b \in \mathcal V^{(a)}})$. Finally, the MTC network contains $R$ MTC servers (or destination nodes) representing the set of destination nodes $\mathcal D$.

The source nodes $\mathcal{S}$, AFC nodes $\mathcal{A}$ and destination nodes $\mathcal D$ are interconnected into an NFC graph $\mathcal{G}=(\mathcal{V}=\mathcal{S}\cup\mathcal{A}\cup\mathcal{D},\mathcal{E}),$ where $\mathcal{V}$ is the set of nodes (modules) and $\mathcal{E} \subseteq {\mathcal V} \times {\mathcal V}$ is the set of edges, i.e., connections between modules. For simplicity, unless otherwise stated, we restrict our attention to directed rooted trees (also called in-trees), where each edge is oriented towards the root node\footnote{We note that this restriction is for simplicity of presentation only; extension to directed acyclic graphs is straightforward and will be required in Sec. V.}. Source nodes $\mathcal{S}$ represent leaves of $\mathcal{G}$. The set of all edges in the graph is completely determined by the set of child nodes of all AFC and destination nodes. We let $\mathcal{V}^{(v)}$ denote the set of child nodes of an arbitrary node $v$. The collection of sets $\{\mathcal{V}^{(v)}\}_{v \in \mathcal{V}}$ fully describes the set of connections between modules.

Finally, we introduce the control elements: \emph{topology processor} and \emph{function processor}, that organize AFC modules into a global NFC evaluator. Based on the MTC server application requirements, the topology and function processors reconfigure the AFC modules to provide a requested MTC-NFC service. In particular, the function processor decomposes a required global NFC into a composition of local AFCs and configures each AFC module accordingly. In other words, based on the requested global network function $f(\mathbf{x}_1,\mathbf{x}_2,\ldots,\mathbf{x}_N)$, the function processor defines a set of atomic functions $\{g^{(a)}\}_{a \in \mathcal{A}}$ and configures the respective AFC modules. Similarly, by defining the graph $\mathcal{G}$ via the set $\{\mathcal{V}^{(v)}\}_{v \in \mathcal{V}}$ and by configuring each AFC node accordingly, the topology processor will interconnect AFC modules into a directed graph of MTC data flows. The topology and function processor are key NFC layer entities. They manage, connect and orchestrate the AFC layer entities, i.e., source modules and AFC modules.

\subsection{CONDENSE Architecture: Implementation}

The above described abstract Condense architecture can be mapped onto the 3GPP MTC architecture. We present initial insights here, while details are left for the following sections.

The AFC layer is composed of AFC modules that evaluate atomic functions. Examples of atomic functions suitable for AFC implementations are the addition, modulo addition, maximum/minimum, norm, histogram, linear combination, threshold functions, etc. Atomic functions can be evaluated straightforwardly in the digital domain using digital processing in network nodes.
In addition to that, atomic functions could be realized by exploiting superposition of signals in the analog domain. Thus, we consider two types of AFC modules: i) Analog-domain AFC (A-AFC), and ii) Digital-domain AFC (D-AFC) modules.

An A-AFC, also referred to as an in-channel AFC, harnesses interference in a wireless channel or signal combining in an optical channel to perform atomic function evaluations. An example of the technology that can be easily integrated as an A-AFC module is the Physical Layer Network Coding~(PLNC) \cite{plnc},~\cite{plncweb}, where the corresponding atomic function is finite field addition, e.g., bit-wise modulo 2 sum in the case of the binary field.

A D-AFC, also referred to as in-node AFC, evaluates atomic functions in the digital domain using, e.g., reconfigurable hardware-based modules in the context of SDN-based implementation \cite{sdnieee}. Alternatively, they can also be implemented using software-based virtual network functions in the context of a NFV-based implementation \cite{Han2015}. An example of the technology that can be easily integrated as a D-AFC module is the packet-level Random Linear Network Coding (RLNC) \cite{fragouli2006},~\cite{practicalNC}. RLNC is a mature technology in terms of optimized software implementations (see, e.g., \cite{kodo}) and it evaluates linear combinations over finite fields as atomic functions. We note that it has been recently proposed and demonstrated within the SDN/NFV framework \cite{rlncsdn},~\cite{rlncnfv}.

The NFC layer can be naturally implemented within the SDN/NFV architecture. In particular, the topology processor naturally fits as an SDN application running on top of the SDN controller within the SDN architecture. In addition, the function processor role may be set within an NFV manager entity, e.g., taking the role of the NFV orchestrator. Using the SDN/NFV framework, MTC-NFC service can be quickly set and flexibly reconfigured according to requests arriving from a diverse set of MTC applications.

\section{Atomic Function Computation Layer}

In this Section, we discuss theoretical and implementation aspects of realizing atomic functions within AFC modules. Subsection~{IV}-A discusses the AFC modules operating in the analog domain, while Subsection~{IV}-B considers digital domain AFCs.

\subsection{Analog-domain Atomic Function Computation (A-AFC)}

\textbf{Wireless-domain A-AFC: Theory}. An A-AFC module's functionality of computing functions over the incoming packets is based on harnessing interference, i.e., the superposition property of wireless channels. We survey the relevant literature on such function computation over wireless channels, finalizing the subsection with presenting current theoretical and technological capabilities.

The idea of harnessing interference for computation is investigated in terms of a joint source-channel communication scheme in \cite{Gastpar2003}, targeting to exploit multiple access channel characteristics to obtain optimal estimation of a target parameter from noisy sensor readings. Extensions of the analog joint source-channel communication are further investigated in the literature, see e.g., \cite{ex1,ex2,ex3,ex4}. Following the impact of network coding ideas across the networking research, reference \cite{plnc} proposes the concept of PLNC to increase throughput of wireless channels; PLNC essentially performs specific A-AFC computations (finite field arithmetics) in a simple two-way relay channel scenario. Computation of linear functions or, more precisely, random linear combinations of the transmitted messages over multiple access channels (MAC) has been considered in \cite{Nazer2007} and extended in \cite{Nazer2011}; therein, the authors propose the compute-and-forward (CF) transmission scheme for computing linear functions at the relays, who attempt to  decode the received random message combinations (the randomness is induced by the fading channel coefficients) to integer combinations, which hence become lattice points in the original code lattice. After this, the relays forward the lattice points to the destination, who can then solve for the original messages provided that the received system of equations is invertible.

Finally, reference~\cite{main} addresses non-linear function computation over wireless channels (see also~\cite{Nazer2007}). While it is intuitive that a linear combination of packets (signals) from multiple sources can be obtained through a direct exploitation of interference, more general, non-linear functions can also be computed through introducing a non-linear (pre-)processing of packets prior to entering the wireless medium, and their (post-)processing after the pre-processed signals have been superimposed in the wireless channel.

\begin{figure}
\centering
\includegraphics[trim = 0mm 65mm 0mm 0mm, clip, width=3.2in]{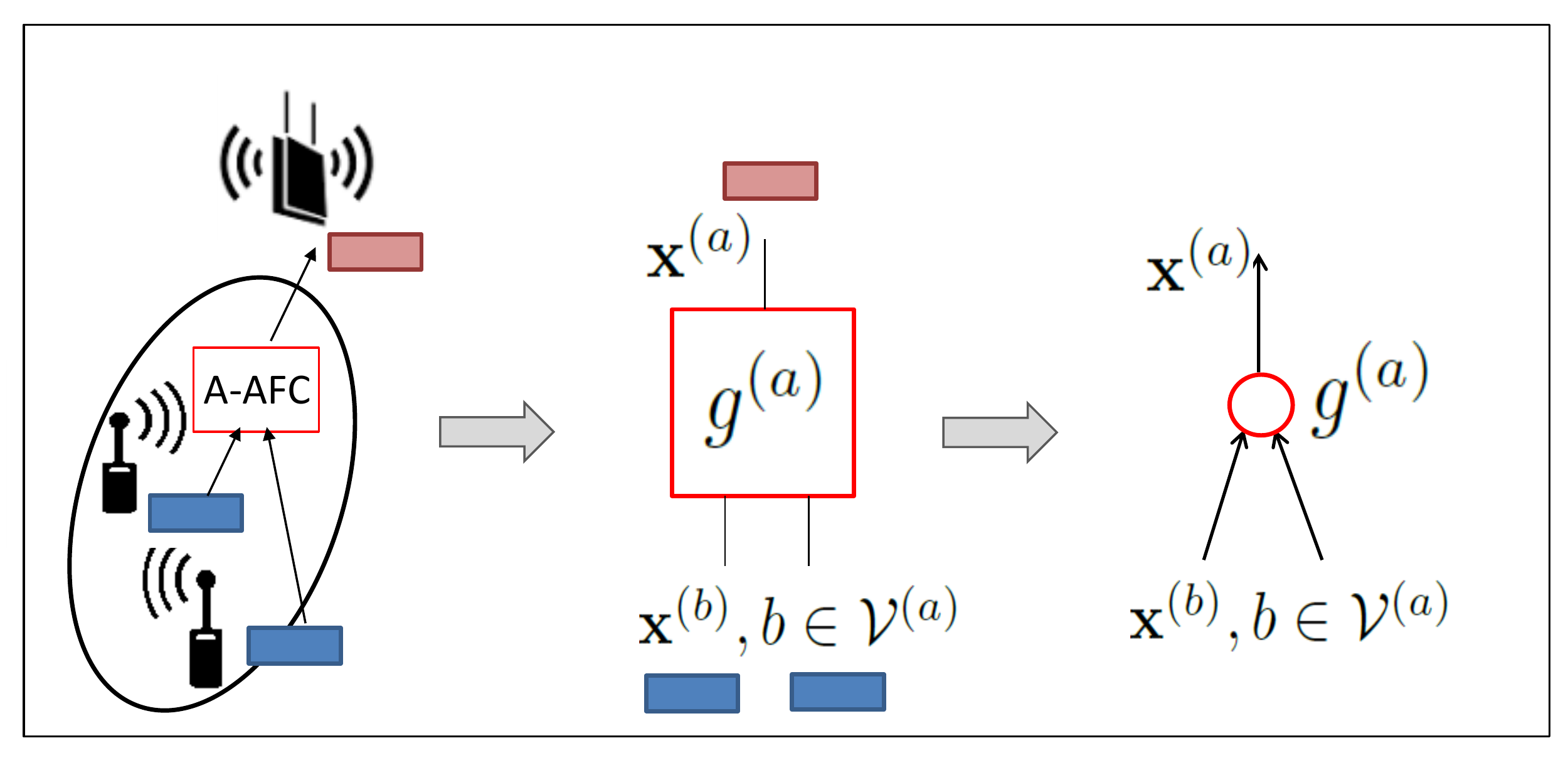}
\caption{Wireless-domain A-AFC module: representation via 3GPP elements (left), modules (middle) and NFC graph nodes (right).}
\label{WD-A-AFC}
\end{figure}

Following~\cite{main}, we now describe in more detail how this non-linear function computation works -- and hence how the A-AFC modules (in principle) operate. Assume that length-$L$ source node data packets $\mathbf{x}_{s}=(x_{s}[1],x_{s}[2],\ldots,x_{s}[L]), s \in \mathcal V^{(a)} \subseteq \mathcal{S}$, arrive at the input interfaces of an AFC node $a$. The packets are first pre-processed by the source node (MTC device) through a pre-processing function $\varphi_s(\mathbf{x}_{s})$. The result is the transmitted symbol sequence $\mathbf{y}_{s}=(y_{s}[1],y_{s}[2],\ldots,y_{s}[L])$, where ${y}_{s}[\ell]=\varphi_s ({x}_{s}[\ell]), \ell=1,2,\ldots,L$. Assuming a block-fading wireless channel model for narrowband signals, the received sequence can be modelled as $\textbf{r}^{(a)}=\left( r^{(a)}[1],r^{(a)}[2],\ldots,r^{(a)}[L] \right)$, where:
\begin{eqnarray}
r^{(a)}[\ell] = \sum_{s \in \mathcal{V}^{(a)}} h_{s} \cdot y_{s}[\ell], \quad \ell=1,2,\ldots,L.
\end{eqnarray}
At the destination, a post-processing function $\psi(\textbf{r}^{(a)})$ is used to obtain $\mathbf{x}^{(a)}$, where $x^{(a)}[\ell] = \psi(r^{(a)}[\ell])$. Therefore, symbol-wise, the A-AFC module~$a$ realizes computation of the following (possibly non-linear) function:
\begin{eqnarray}
\label{eqn-2-nomog}
g^{(a)}\left( \{x_s[\ell]\}_{s \in \mathcal{V}^{(a)}}\right)
 = \psi\left( \sum_{s \in \mathcal{V}^{(a)}} h_{s} \,\varphi_s ({x}_{s}[\ell]) \right).
\end{eqnarray}
The class of functions computable via A-AFC modules, i.e., which are of form~\eqref{eqn-2-nomog}, are called nomographic, and they include important functions such as the arithmetic mean and the Euclidean norm~\cite{main}.

Fig. \ref{WD-A-AFC} illustrates an A-AFC: its position in the real-world system (left), its representation as an A-AFC module (central), and as part of the NFC graph (right).   

\begin{figure}
\centering
\includegraphics[trim = 0mm 55mm 80mm 0mm, clip, width=3.2in]{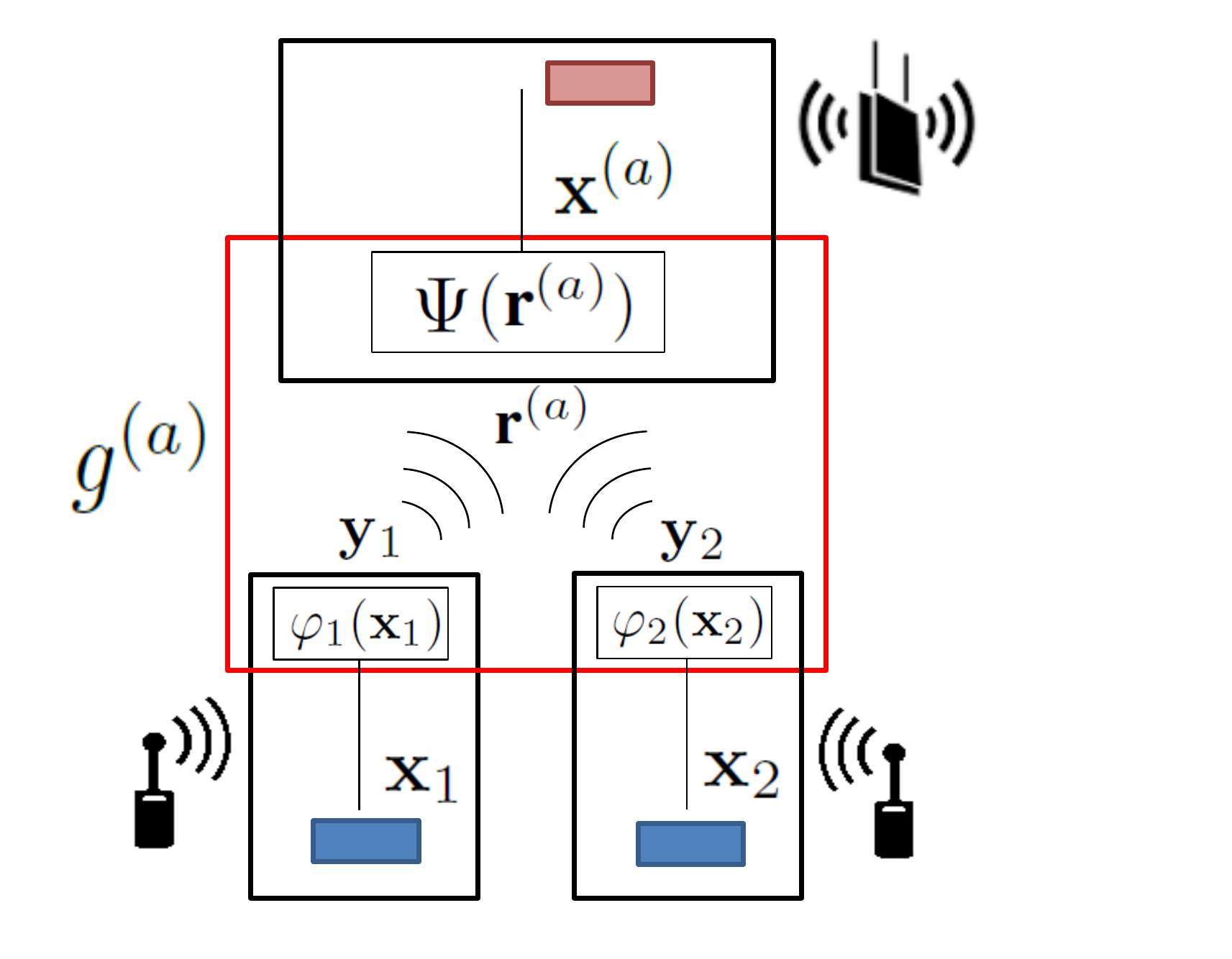}
\caption{Wireless-domain A-AFC module: Functional diagram.}
\label{WD-A-AFC-2}
\end{figure}

\textbf{Wireless-domain A-AFC: Implementation}. The above framework implies that an A-AFC module is physically spread across all input devices and the output device connected to the A-AFC module, as illustrated in Fig. \ref{WD-A-AFC-2}. At the input devices (e.g., MTC devices), an appropriate input A-AFC digital interface needs to be defined that accepts input data packets and implements pre-processing function $\varphi(\cdot)$ before the signal is transmitted into the channel. Similarly, at the output device, e.g., small base station (HeNB), an appropriate output A-AFC digital interface needs to be defined that delivers output data packets after the signal received from the channel is post-processed using $\psi(\cdot)$. We also note that, although above we assume input nodes to A-AFC module are source nodes (MTC devices), wireless-domain A-AFC module can be part of the wireless backhaul network, e.g., connecting several HeNBs to the eNB.

\textbf{Challenges and Future Directions}. In the current state-of-the-art, A-AFC is investigated in the context of joint computation and communications in wireless sensor networks. Current research works are limited in terms of the computed functions, such as addition, multiplication, norm, arithmetic/geometric means, and are also limited in scope, as they are only targeting the wireless -- and not optical -- communication links. Design and implementation of generic A-AFC in wireless setting which is adaptive to the channel conditions remains an open problem. Note that any such design should take practical implementation aspects into account, including channel estimation errors, timing and frequency offsets and quantization issues. Furthermore, the link qualities between network nodes, including adaptive schemes that select the computation nodes according to the robustness of communications between links, need to be considered to improve the reliability of the computed function outputs.

\textbf{Optical-domain A-AFC: Discussion}. If we consider the PLNC example, it is clear that wireless domain A-AFC modules are close to become a commercially available technology (see, e.g., \cite{plncweb}). The question that naturally arises is whether A-AFC modules can be implemented in optical channels within optical access networks such as passive optical networks (PON). This would further increase the richness of AFC layer and bring novel AFC modules into the MTC-NFC network. Here, we briefly comment on the status of optical-domain function computation.

Recent works analyzed applicability of network coding of data packets within PONs in some simple scenarios \cite{Miller2010}\cite{NCPON}. However, in contrast to the above vision of A-AFC modules, in these works signals are not ``in-channel'' combined, rather, network coding is done at the end-nodes, in the digital domain.

Information processing in the photonic domain has been envisioned in the 1970s. But implementations of digital optical computing could not keep pace with the development of electronic computing. Nevertheless, with advances in technology, the role of optics in advanced computing has been receiving reawakened interest \cite{Caulfield}. Moreover, unconventional computing techniques, in particular reservoir computing (RC), find more and more interest and are being implemented in different photonic hardware. RC is a neuro-inspired concept for designing, learning, and analysing recurrent neural networks – neural networks where, unlike the most popular feed-forward neural networks, the interconnection network of neurons possesses cycles (feedback loops). A consequence of the presence of loops is, as pointed out in \cite{Lukosevicius}, that recurrent neural networks can process and account for temporal information at their input. A recent breakthrough was a drastic simplification of the information-processing concept of reservoir computing (RC) in terms of hardware requirements \cite{Appeltant}. The appeal of RC therefore resides not only in its simple learning, but moreover in the fact that it enables simple hardware implementations. Complex networks can be replaced by a single or a few photonic hardware nodes with delayed feedback loops \cite{Larger2012}, \cite{Paquot2012}, \cite{Brunner}. Different tasks, including spoken digit recognition, nonlinear time series prediction and channel equalization have been performed with excellent performance, speed and high energy efficiency \cite{Brunner}, \cite{Vandoorne}. Beyond these first successes, meanwhile, using simple hardware, learning approaches including RC, extreme learning machines and back-propagation learning of recurrent neural networks have been demonstrated \cite{Hermans2015}, illustrating the flexibility and potential of this approach.

\subsection{Digital-domain Atomic Function Computation (D-AFC)}

\begin{figure}
\centering
\includegraphics[trim = 0mm 55mm 0mm 0mm, clip, width=3.2in]{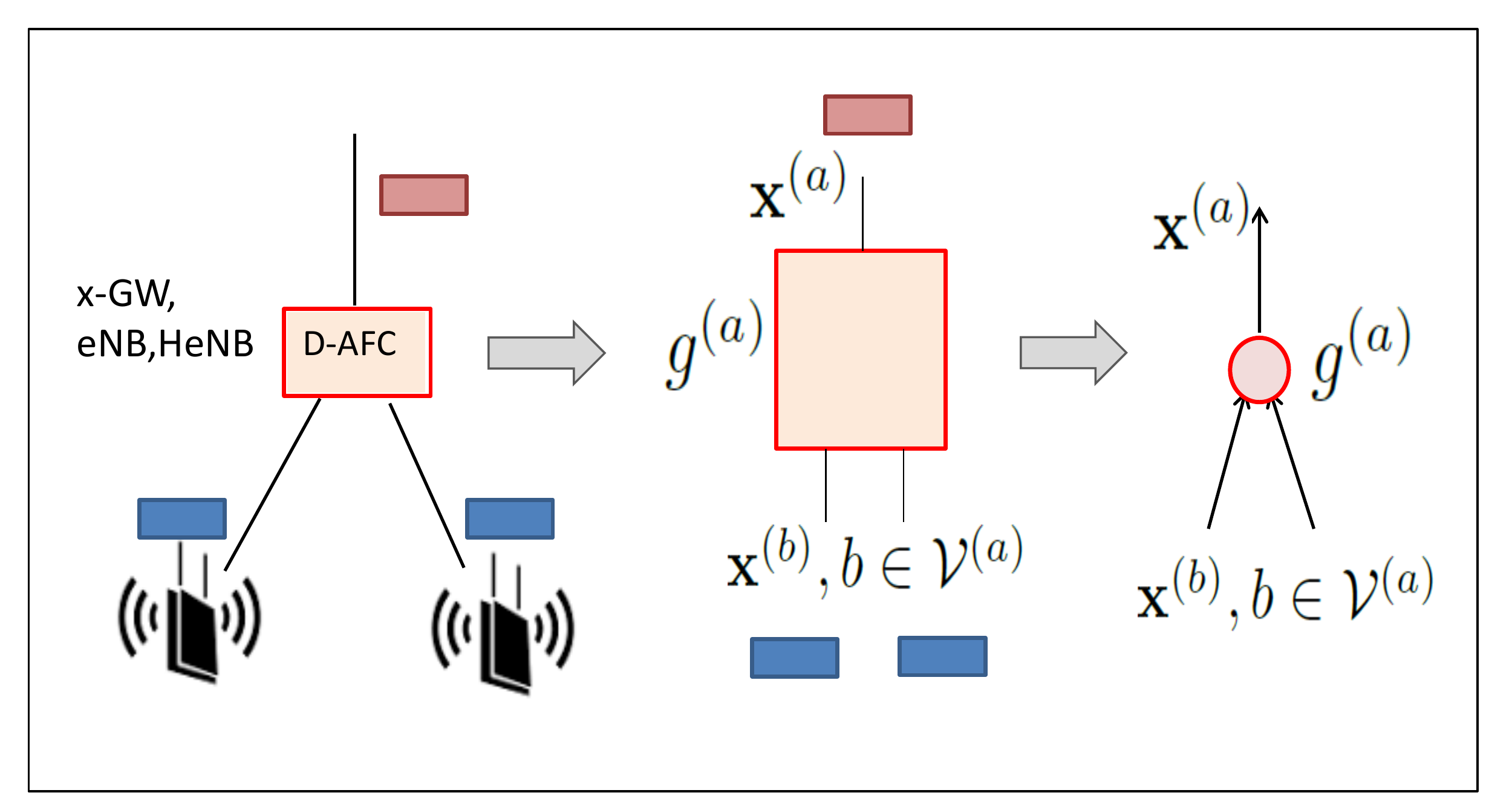}
\caption{D-AFC module: representation via 3GPP elements (left), modules (middle) and NFC graph nodes (right).}
\label{D-AFC}
\end{figure}

D-AFC modules evaluate atomic functions in the digital domain, within the network nodes such as base stations (eNB or HeNB) and core network gateways (HeNB-GW, S-GW, P-GW). Although digital-domain in-node processing offers many possibilities for D-AFC implementation, here we address two possible options suitable for the SDN and NFV architectures.

The first option for D-AFC are reconfigurable hardware-based Field Programmable Gate Array (FPGA) platforms. FPGA platforms are frequently used in combination with high-speed networking equipment to perform various work-intensive and high-throughput demanding functions over data packets such as packet filtering \cite{netfpga}. FPGAs are either integrated in network nodes as co-processing units, or can be easily attached as external units to network nodes via high-speed network interfaces. FPGAs offer flexible and reconfigurable high-throughput implementations of various linear or non-linear atomic functions. For example, implementing random linear combinations over input data packets in network nodes -- as part of RLNC -- is considered in several recent works \cite{rlncfpga1},~\cite{rlncfpga2}. D-AFC implementations via FPGA platforms offer seamless integration in SDN concepts, because SDN data flows can be easily filtered and fed into either internal or external FPGA units. Depending on the application, FPGAs achieve speed increase over general processing units by factor of tens to hundreds. Note also that FPGAs can be reprogrammed and reconfigured in short time intervals (order of minutes).

The second possibility for efficient D-AFC is to use software-based implementations in high-level programming languages that run on general processing units, either in network nodes or externally on dedicated general-purpose servers \cite{kodo}. This approach offers full flexibility for atomic function evaluation at the price of lower data processing throughput as compared with the FPGA approach. An example of a D-AFC implementation of random linear combinations over incoming data packets in the context of RLNC is given in \cite{rlncsdn},~\cite{rlncnfv}. Software-based D-AFC implementations can be easily and remotely instantiated across the network nodes in a virtualized environment following NFV concepts.

Fig. \ref{D-AFC} illustrates a D-AFC: its position in the real-world system (left), its representation as an D-AFC module (center), and as part of the NFC graph (right).

\section{Network Function Computation Layer}

The NFC layer is responsible for configuring the Condense topology and assigning the appropriate atomic functions across the AFC modules, such that a desired network-wide function computation is realized. Subsection~{V-A} discusses theoretical aspects (capabilities and limitations) of computing functions over networks, surveying the relevant literature on sensor fusion and network coding for computing. Subsection~{V-B} describes a possible implementation of NFC functionalities within the 3GPP MTC system, through a more detailed view of SDN/NFV modules, i.e., the function and topology processors.

\subsection{Theoretical Aspects of NFC Layer}


The need for mathematical theory of function computation in networks is advocated in~\cite{Giridhar05},~\cite{Giridhar}. The authors discuss various challenges in sensor networks, and argue that computation of functions in a sensor network could lead to a lower data overhead, as well as to a reduced data traffic. For our toy example, in the fire alarm sensor network, we are only interested in the measurements of the highest temperature in the set of sensors. Alternatively, in monitoring temperature range in a green house, we might only be interested in the measurements of the average temperature from the set of sensors. Therefore, for various practical applications, it would be beneficial if the network node would be able to perform basic (atomic) computation, which in the context of the whole network could lead to computation of more sophisticated functions in the destination nodes.

This subsection elaborates on the mathematical tools behind the realization of the Condense NFC layer. There is a number of works studying function computation over a network, which are available in the literature. The relevant work includes those in contexts of sensor fusion, network coding for computing, and neural networks. The two former work threads are discussed here, while the latter is discussed in Subsection~{VI-C}. Hereafter, we mostly follow the framework defined in~\cite{AFKZ,Kowshik2012}, adapting notation to our needs here.

\textbf{Mathematical settings}. Consider a finite directed acyclic graph $\mathcal G = (\mathcal V, \mathcal E)$, consisting of $M$ AFC nodes belonging to set $\mathcal{A}$, a set of $N$ sources (MTC devices) $\mathcal{S}$, and a set of $R$ destinations $\mathcal{D}$, such that $\mathcal{S} \cap \mathcal{D} = \varnothing$. 

The network uses a finite alphabet $\mathbb A$, called \emph{network alphabet}. Each source $s$ generates $K$ random symbols
$\sigma_s[1], \sigma_s[2],\ldots,\sigma_s[K] \in \mathbb A$. Here, we say that the source
 symbol $\sigma_s[k]$ belongs to the $k$-th generation of the source symbols.

We assume that each packet sent over a network link is a vector of length $L$ over $\mathbb A$. Suppose that each of the $R$ destination nodes requests computation of a (vector-valued) function $f$ of the incoming MTC device
vectors ${\boldsymbol \sigma}_s,$ $s=1,...,N$. The target vector function is of the form 
$f: {\mathbb A}^{N\cdot K} \rightarrow {\mathbb B}^K$, where $\mathbb B $ is a function alphabet, 
and each component function $f: {\mathbb A}^{N} \rightarrow {\mathbb B}$ is of the same form, applied to each 
source's $k$-th symbol, $k=1,...,K$. More precisely, we wish to compute 
$f \left( \sigma_1[k],...,\sigma_N[k]\right)$, $k=1,...,K$.


With each arc $a \rightarrow v$ outgoing an AFC node~$a \in \mathcal{A}$, we associate the atomic function $g^{(a \rightarrow v)} \left( \cdot \right)$, which takes
the~$|\mathcal{V}^{(a)}|$ length-$L$ incoming vectors $\mathbf{x}^{(u \rightarrow a)}$, $u \in {\mathcal V}^{(a)}$,
and produces the length-$L$ outgoing vector~$\mathbf{x}^{(a \rightarrow v)}$, i.e.:
 \[
 \mathbf{x}^{(a \rightarrow v)} = g^{(a \rightarrow v)} \left( \{\mathbf{x}^{(u \rightarrow a)}\}_{u \in \mathcal{V}^{(a)}}\right).
 \]

Similarly, with each arc $s \rightarrow v$ outgoing a source node~$s \in \mathcal{S}$, the atomic function $g^{(s \rightarrow v)} \left( \cdot \right)$  takes the~$|\mathcal{V}^{(s)}|$ length-$L$ incoming vectors $\mathbf{x}^{(u \rightarrow s)}$, $u \in {\mathcal V}^{(s)}$, as well as the $K$ generated symbols $\boldsymbol{\sigma}_s$ = $(\sigma_s[1],...,\sigma_s[K])$, and produces the length-$L$ outgoing vector~$\mathbf{x}^{(s \rightarrow v)}$, i.e.:
 \[
 \mathbf{x}^{(s \rightarrow v)} = g^{(s \rightarrow v)} \left( \{\mathbf{x}^{(u \rightarrow s)}\}_{u \in \mathcal{V}^{(s)}};\,\boldsymbol{\sigma}_s\right).
 \]
(Note that we consider here the most general case in which a source node does not have to lie on the ``bottom-most'' level of the network, i.e., it can also have some incoming edges.) We refer here to both $g^{(a \rightarrow v)}$'s and $g^{(s \rightarrow v)}$'s as \emph{encoding functions}.

\begin{figure*}
\centering
\includegraphics[trim = 0mm 20mm 0mm 0mm, clip, width=6.6in]{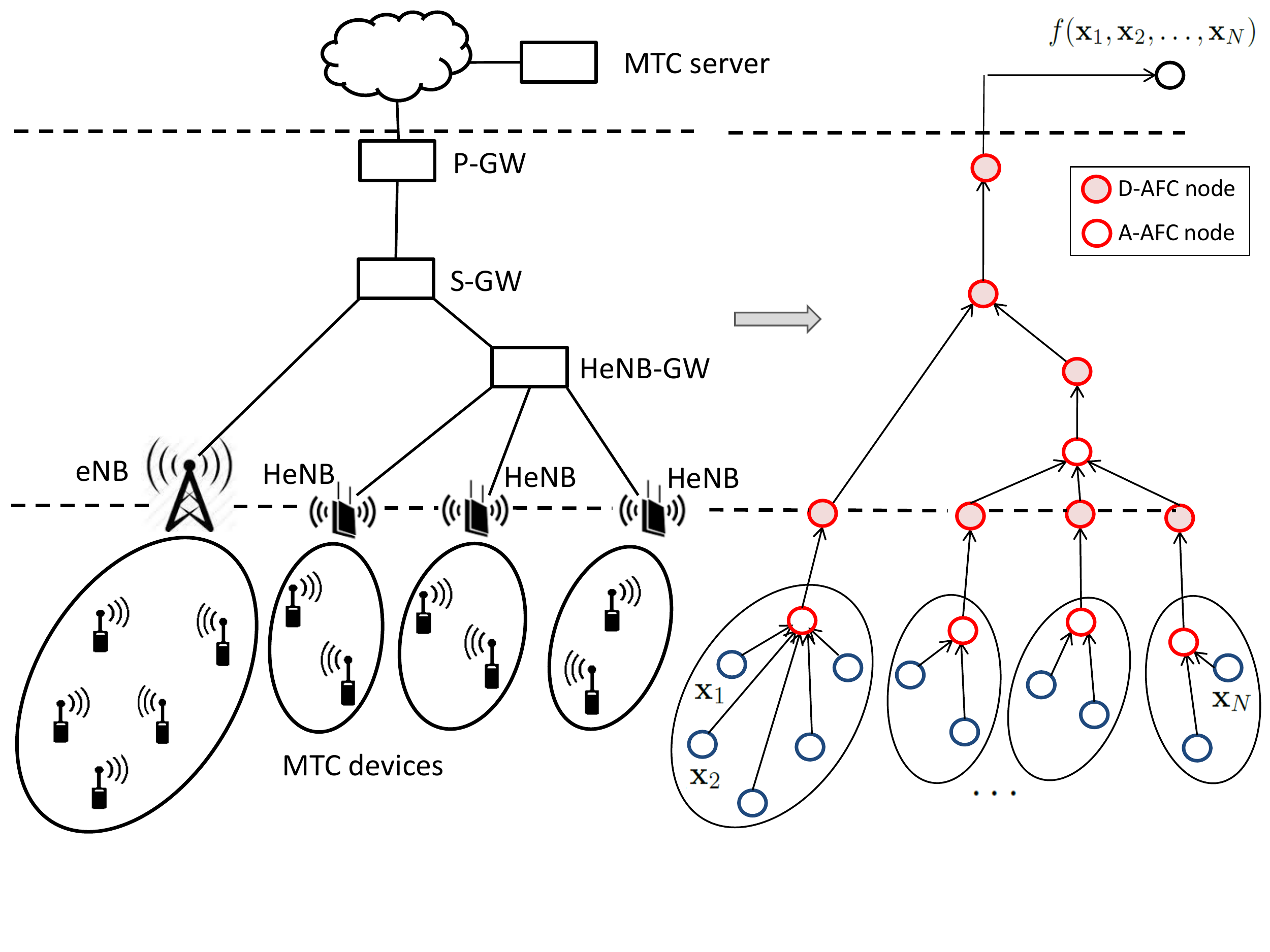}
\caption{Mapping between MTC network and NFC graph.}
\label{NFCgraph}
\end{figure*}

Finally, a destination node $d \in \mathcal D$ takes its~$|\mathcal{V}^{(d)}|$ incoming length-$L$ messages and performs decoding, i.e., it produces the vector of function evaluation estimates $\widehat{\mathbf{f}}^{(d)} = \left(\widehat{f}^{(d)}[1],...,\widehat{f}^{(d)}[K]\right)$, as follows:
 \[\widehat{\mathbf{f}}^{(d)} = \Psi^{(d)}\left( \{\mathbf{x}^{(u \rightarrow d)}\}_{u \in \mathcal{V}^{(d)}}\right),\]
where $\Psi^{(d)}(\cdot)$ is the destination node $d$'s function. Note that $\Psi^{(d)}(\cdot)$ recovers back the 
$K$-dimensional vector from the $L$-dimensional incoming quantities (where $L>K$), and it is therefore referred to as a decoding function.

We say that the destination $d \in \mathcal D$ \emph{computes} the function $f \; : \; {\mathbb A}^N \rightarrow {\mathbb B}$, if for every generation $k \in \{1,...,K\}$, it holds that:
\[\widehat{f}^{(d)}[k] = f \left( \sigma_1[k],...,\sigma_N[k]\right).\]
Further, we say that the problem of computing $f$ is \emph{solvable} if there exist atomic functions $g^{(s \rightarrow v)}\left( \cdot\right)$, $g^{(a \rightarrow v)}\left( \cdot\right)$ across all arcs in $\mathcal E$ and decoding functions $\Psi^{(d)} \left( \cdot \right)$, $d =1,...,R$, such that $f$ is computed at all destinations $d \in \mathcal D$ (that is, their corresponding composition computes $f$ at all destinations).

\textbf{Connection to network coding and beyond}.
The reader can observe that the problem of network coding~\cite{Ahlswede} is a special case of the function computation problem with $L = K = 1$, where the target function $f$ is an identity function: $f(\sigma_1, \sigma_2, \cdots, \sigma_N ) = (\sigma_1, \sigma_2, \cdots, \sigma_N ).$ In particular, in linear network coding, the alphabet $\mathbb A$ is taken as a finite field $\mathbb F$, the function alphabet $\mathbb B$ is $\ff^N$,
and all encoding functions $g^{(a \rightarrow v)}$, $g^{(s \rightarrow v)}$ and decoding functions $\Psi^{(d)}$ are linear mappings over $\ff$. The case of linear network coding is relatively well understood. In particular, it is known that the problem of computing $f$ is solvable if and only if each of the minimum cuts between all the sources and any destination has capacity of at least $N$~\cite{KM}. In Subsection~{VI-A}, we provide further details on this special case.

For non-linear network coding, the universal criteria for network coding problem solvability are not fully understood. It is known, for example, that for the case where each sink requests a subset of the original messages, there exist networks,
which are not solvable by using linear functions  $g^{(a \rightarrow v)}(\cdot)$, $g^{(s \rightarrow v)}(\cdot)$ and $\Psi^{(d)}(\cdot)$, yet they can be solved by using non-linear functions (see, for example,~\cite{Dougherty}).

In order to understand the fundamental limits on solvability of the general function computation problem, the authors of~\cite{AFKZ} define what they term the computing capacity of a network as follows:
\begin{multline}
\cc(\mathcal G,f) = \\
\sup \left\{ \frac{K}{L} \; : \; \mbox{computing $f$ in $\mathcal G$ is solvable } \right\} \; .
\end{multline}
They derive a general min-cut type upper bound on the computing capacity, as well as a number of more specific lower bounds. In particular, special classes of functions, such as symmetric functions, divisible functions and exponential functions, are considered therein (see~\cite{Dougherty} for more detail). It should be mentioned that the considered classes of functions are rather restricted, and that they possess various symmetry properties. The problem turns
out to be very difficult, however, for more general, i.e., less restricted, classes of functions.

Another related work is~\cite{AF}, where a set-up with linear functions $g^{(a \rightarrow v)}(\cdot)$, $g^{(s \rightarrow v)}(\cdot)$ and $\Psi^{(d)}(\cdot)$ and general linear target function $f$ is considered. The authors are able to characterize some classes of functions, for which the cut-set bound gives sufficient condition for solvability, and for which it does not.

\textbf{Other results}. In~\cite{Giridhar05}, the function computation rate is defined and lower bounds on such rate are obtained for various simple classes of functions. Recently, in~\cite{ST}, information-theoretic bounds on the function computation rate were obtained for a special case, when network is a directed rooted tree, and the set of sinks contains only the root of the tree, and the source symbols satisfy a certain Markov criterion.

A number of works study computation of sum in the network. It is shown in~\cite{RaiDey} that if each sink requests a sum of the source symbols, the linear coding may not be sufficient in networks where non-linear coding can be sufficient. Other related works include~\cite{RL,Kannan, Kowshik, Shah, Lalitha}.

%

There is a significant number of works related to \emph{secure} function computation available in the literature. However, usually, the main focus of these works is different. We leave that topic outside of the scope of this paper.

\textbf{Challenges and research directions}. Research on network function computation is still in its infancy and general theoretic foundations are yet to be developed. Here, we identify several challenges with network function computation relevant for Condense architecture. First, it is important to consider the issue of solvability when the encoding and decoding functions $g^{(a \rightarrow v)}(\cdot)$, $g^{(s \rightarrow v)}(\cdot)$ and $\Psi^{(d)}(\cdot)$ are restricted to certain classes dictated by the underlying physical domain. For instance, A-AFC modules operating in the wireless domain are currently restricted to a certain class of functions (see Subsection~{IV-A}), while, clearly, D-AFCs operating in the digital domain have significantly more powerful capabilities. Second, it interesting to study NFC in simpler cases, when the network topology is restricted to special classes of graphs, for example rooted trees, directed forests, rings, and others. Third, under the above defined constraints on the AFC capabilities, the question that arises is how well we can approximate a desired function, even if solvability is impossible. Finally, practical and efficient ways for \emph{actual constructions} of $g^{(a \rightarrow v)}(\cdot)$, $g^{(s \rightarrow v)}(\cdot)$ and $\Psi^{(d)}(\cdot)$, as opposed to existence-type results, are fundamental for Condense implementation.

\subsection{Implementation Aspects of NFC Layer}

In practical terms, the NFC layer should deal with control and management tasks of establishing and maintaining an NFC graph of AFC modules for a given service request, as sketched in Fig. \ref{NFCgraph}. In our vision, the main control modules that define the NFC layer functionality are: the function processor (FP) and the topology processor (FP) (see Fig. \ref{MTC-NFC}). Both modules can be seamlessly integrated in the SDN/NFV architecture.

The TP module organizes the MTC data flows and sends configuration instructions via the SDN control plane. In abstract terms, for all nodes in a directed rooted tree (or directed acyclic graph), TP needs to provide the set of child nodes $\{\mathcal{V}^{(v)}\}_{v \in \mathcal{V}}$ from which to accept MTC data flows, and to identify the exact MTC data flows that will be filtered for each output flow, if there are multiple output flows.

Based on the MTC server application requests and the configured topology, FP processes the global function request, and, based on the available library of AFC modules, it generates the set of atomic functions to be used: $\{g^{(a)}\}_{a \in \mathcal{A}}$. Note that, as described before, AFC modules may be: i) A-AFC modules, ii) hardware-based D-AFC modules, and ii) software-based D-AFC modules. A-AFC modules (e.g., PLNC module) need more complex instantiation control as they spread over several physical nodes and involve configuration of input and output interfaces and pre/post-processing functions (Sec. IVA). Hardware-based D-AFC modules (e.g., internal/external FPGA modules within or attached to network elements) require SDN-based control of MTC data flows that will filter selected flows and direct them through the D-AFC module. Finally, the most flexible case of software-based D-AFC modules (e.g., software-based modules in virtual machines running over the virtualized hardware in network elements or external servers) is a library of AFC implementations where each atomic function from the library can be remotely instantiated via the NFV control. Overall, FP needs to know the list of available AFC resources in the entire NFC network in order to optimize the set of instantiated AFC modules.

\begin{figure}
\centering
\includegraphics[trim = 0mm 45mm 10mm 0mm, clip, width=3.2in]{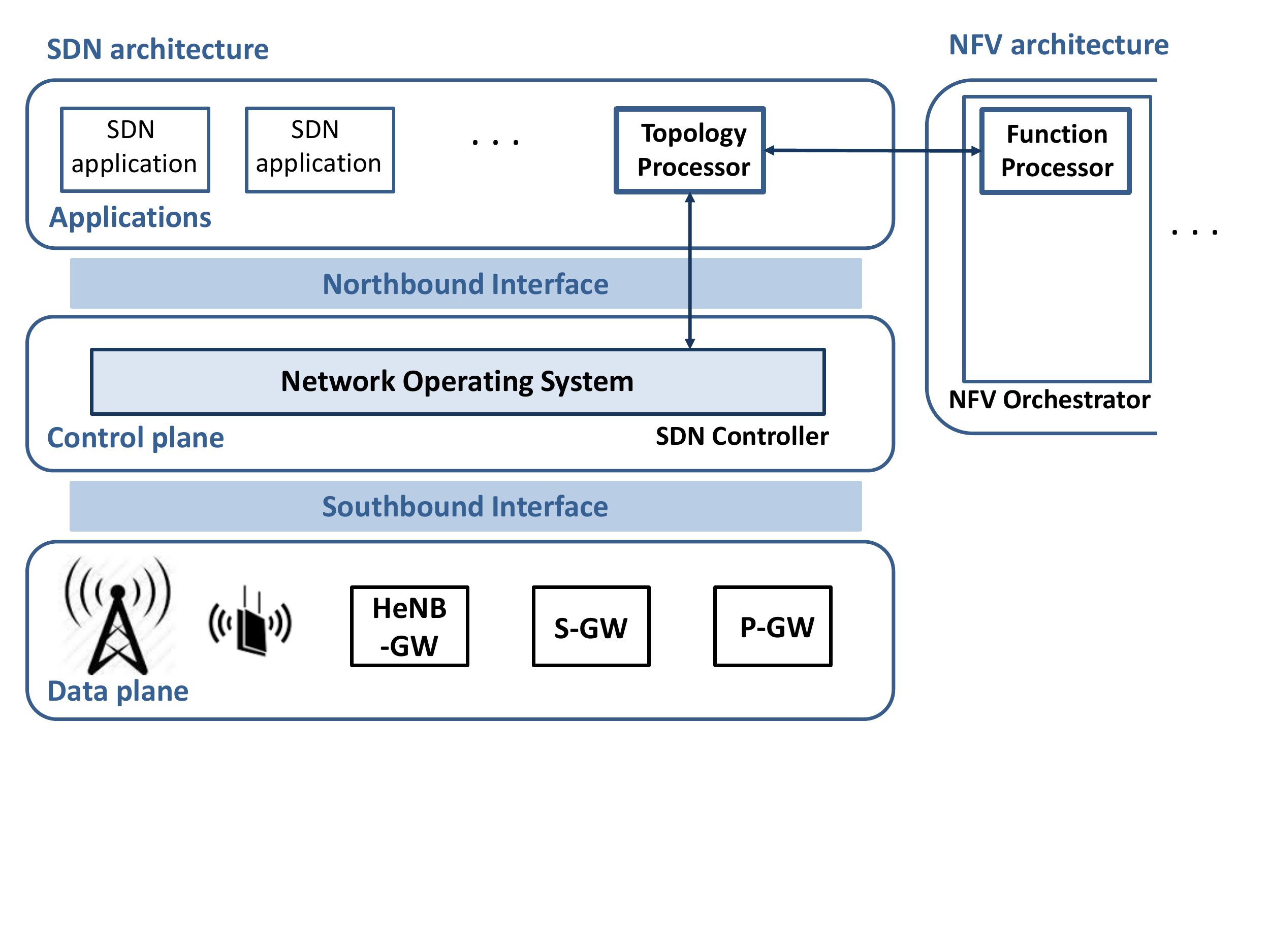}
\caption{NFC layer modules within SDN/NFV architecture.}
\label{NFC-modules}
\end{figure}

In terms of realization, we use the standard proposal for NFV/SDN complementary coexistence \cite{Li2015} where the FP module can be implemented as an NFV architecture block called the NFV orchestrator. The TP module can be implemented as an SDN application. Besides communicating directly, both FP and TP modules, observed as SDN applications, approach the SDN control plane via the SDN northbound interface. Based on the FP/TP inputs, the SDN control plane will configure physical devices (network nodes) via the southbound interface. This NFV/SDN based control of Condense is illustrated in Fig.~{7}.

\section{Application Layer}

In this section, we present three examples of applications at the application layer of the Condense architecture: data recovery through network coding, minimizing population risk via a stochastic gradient method, and binary classification via neural networks\footnote{Strictly speaking, learning a neural network can be considered a special case of a stochastic gradient method (with a non-convex loss function). We present it here as a distinct subsection as we consider the implementation where the neural network weight parameters are distributed across the Condense network.}. The purpose of these examples is two-fold. First, they demonstrate that a wide range of applications can be handled via the Condense architecture. Second, they show that Condense is compatible with widely-adopted concepts in learning and communications, such as random linear network coding, stochastic gradient methods, and neural networks.

The three examples are also complementary from the perspective of the workload required by the FP module. With the first example (data recovery via network coding), the function of interest is decomposed into mutually uncoordinated atomic (random) linear functions, and hence no central intervention by the function controller is required, nor is the inter-AFC modules coordination needed as long as atomic functions are concerned. With the third example (binary classification via neural networks), the desired network-wide function is realized through a distributed coordination of the involved AFC modules. Finally, with the second example (minimizing population risk via a stochastic gradient method), the most generic case requires the intervention of the central FP module, in order that the desired network-wide function be decomposed and computed.

\subsection{Data recovery through network coding}

A special case of an application task with the Condense architecture is to deliver the \emph{raw data} to the data center of interest. This corresponds to a trivial, identity function over the input data as a goal of the overall network function computation. However, this is not achieved through simply forwarding the raw data to the data center, but through the usage of network coding. In other words, atomic functions are not identity functions but random linear combinations over the input data. While such solution may not reduce the total communication cost with respect to the conventional (forwarding) solution, this solution is significantly more flexible, robust and reliable, e.g.,~\cite{fragouli2006},~\cite{practicalNC}. As recently noted, it can be flexibly implemented within the context of network coded cloud storage \cite{RLNCcloud}.

We follow the standard presentation of linear network coding, e.g.,~\cite{fragoulibook}, adapting it to our setting. For ease of presentation, we assume here that graph $\mathcal G$ is a directed rooted tree. Therein, the destination node is the root of the tree. Suppose that each of the $N$ available MTC devices has a packet $\mathbf{x}_s$ consisting of $L$ symbols, each symbol belonging to a finite field~$\mathbb F$. We adopt the finite field framework as it is typical with network coding. The goal is to deliver the whole packet vector $\mathbf{x}=(\mathbf{x}_1,...,\mathbf{x}_N)$ to the data center (the destination node~$d$). With Condense, this is achieved as follows. Each atomic node~$a$ generates the message pair $\left( \mathbf{x}^{(a)}, \mathbf{c}^{(a)} \right)$ (to be sent to the parent node) based on the received messages from its child nodes $\left( \mathbf{x}^{(b)}, \mathbf{c}^{(b)} \right)$, where $b \in {\mathcal V}^{(a)}$, and we recall that $\mathcal{V}^{(a)}$ is the set of child nodes of node~$a$. As we will see, the quantity $\mathbf{x}^{(a)} \in {\mathbb F}^L$ is by construction a linear combination of the (subset of) MTC devices' packets $\mathbf{x}_s \in {\mathbb F}^L$, $s=1,...,N$. The quantity $\mathbf{c}^{(a)} = (c^{(a)}[1],...,c^{(a)}[N]) \in {\mathbb F}^N$ stacks the corresponding weighting (or coding) coefficients; that is:
     \begin{equation}
     \label{eqn-net-cod-1}
     \mathbf{x}^{(a)} = \sum_{s=1}^N c^{(a)}[s]\,\,\mathbf{x}_s.
     \end{equation}
Now, having received $\left( \mathbf{x}^{(b)}, \mathbf{c}^{(b)} \right)$, $b \in {\mathcal V}^{(a)}$, node $a$ computes $\mathbf{x}^{(a)}$ using random linear network coding approach. It first generates new random (local) coding coefficients ${e}^{(a)}[b] \in \mathbb F$, $b \in {\mathcal V}^{(a)}$, uniformly from $\mathbb F$, and independently of the received messages. Then, it forms $x^{(a)} $ as:
      \[
     x^{(a)} = \sum_{b \in \mathcal{V}^{(a)}} e^{(a)}[b] \,\,\mathbf{x}^{(b)}.
     \]
Once $\mathbf{x}^{(a)}$ has been computed, node $a$ also has to compute the global coding coefficients $\mathbf{c}^{(a)}$ \emph{with respect to the MTC packets} $\mathbf{x}_s$, $s=1,...,N$, as per~\eqref{eqn-net-cod-1}. It can be shown that:
     \[
     c^{(a)}[s] = \sum_{b \in \mathcal{V}^{(a)}} e^{(a)}[b] \,\,c^{(b)}[s],\,s=1,...,N.
     \]
For the end-leaf (MTC device) nodes~$s$, we clearly have that $c^{(s)}[s]=1,$ and $c^{(s)}[u] =0$, $u \neq s$.
Once the destination (root) node~$d$ receives all its incoming messages, it has available a random linear
combination of the MTC's packets $\mathbf{x}_1,...,\mathbf{x}_N$:
       \[
       \mathbf{x}^{(d),1} = \sum_{s=1}^N c^{(d),1}[s]\,\,\mathbf{x}_s,
       \]
and the corresponding global coding coefficients vector~$\mathbf{c}^{(d),1} = \left( c^{(d),1}[1],...,c^{(d),1}[N] \right)$.        Afterwards, the whole process described above is repeated sequentially $N^\prime -1$ times, such that the data center obtains~$N^\prime-1$ additional pairs $\left( \mathbf{x}^{(d),k}, \mathbf{c}^{(d),k}\right)$, $k=2,...,N^\prime$. It can be shown that, as long as $N^\prime$ is slightly larger than~$N$, MTC data vector $\mathbf{x}=\left(\mathbf{x}_1,...,\mathbf{x}_N\right)$ can be recovered with high probability through solving the linear system
 of equations with unknowns~$\mathbf{x}_s$:
          \[
          \mathbf{x}^{(d),k} = \sum_{s=1}^N c^{(d),k}[s]\,\,\mathbf{x}_s,\,k=1,...,N^\prime.
          \]
Note that, for this application example, each atomic function is \emph{linear}. Moreover, there is no requirement on the \emph{coordination} of the atomic functions which correspond to different atomic nodes, as they are generated randomly and mutually independently \cite{Ho2006}. Hence, this application does not require a centralized control by the FP module. 

Finally, when certain a priori knowledge on $\mathbf{x}$ is available (e.g., sparsity, i.e., many of the packets $\mathbf{x}_s$ are the zero $L$-tuples of symbols from $\mathbb F$), then the recovery probability close to one can be achieved even when the number of linear combinations $N^\prime$ at the MTC server is significantly smaller than~$N$. Omitting details, this can be in principle achieved using the theories of compressed sensing and sparse recovery, e.g., \cite{Candes},~\cite{Hayashi}.

\subsection{Statistical estimation and learning}
A dominant trend in current machine learning research are algorithms that scale to large datasets and are amenable to modern distributed processing systems. Machine learning systems are widely deployed in architectures with a large number of processing units at different physical locations and communication is becoming a resource that is taking the center stage in the algorithm design considerations \cite{Jordan2015,Zhang2013,Agarwal2011,Boyd2011,Shamir2014}.

Typically, the task of interest (parameter estimation, prediction, etc.) is performed through solving an optimization problem of minimizing a risk function \cite{Wasserman2004}. In most widely used models of interest, which include logistic regression and neural networks, this optimization needs to be performed numerically using gradient descent methods and is simply based on successive computation of the gradient of the loss function of interest. In large datasets (of size $T$), obtaining the full gradient comes with a prohibitive computational (a linear computational cost in $T$ \emph{per iteration} of gradient descent cannot be afforded) as well as a prohibitive communication cost (due to the need to access \emph{all} training examples even though they may be, and typically are, stored at different physical locations). For these reasons, stochastic gradient methods are the norm -- they typically access only a small number of data points at a time, giving an unbiased estimate to the gradient of the loss function needed to update the parameter values -- and have enjoyed tremendous popularity and success in practice \cite{Bottou2011,Yousefian2012,NiuArxiv,Bottou2010}.

Most existing works assume that the data has already been collected and transmitted through the communication architecture and is available at request. That is, typically the data is first transmitted in its raw form from the MTC devices to the data center, and only afterwards a learning (optimization) algorithm is executed.

In contrast, the Condense architecture integrates the learning task into the communication infrastructure. That is, a learning task is seen as a sequence of oracle calls to a certain network function computation, and the role of the NFC layer is to provide these function computations at the data center's processing unit (destination node) and only the computed value (e.g., of the gradient to the loss function) is being communicated.  This way, Condense will generically embed various learning algorithms into the actual 3GPP MTC communication infrastructure.

We now dive into more details and exemplify learning over the proposed Condense system with the estimation of an unknown parameter vector $\mathbf{w}^\star \in {\mathbb R}^Q$ through the minimization of population risk. Specifically, we consider stochastic gradient-type methods to minimize the risk.

To begin, consider a directed rooted tree NFC graph $\mathcal G$, and assume that there are $N$ MTC devices which generate samples $\mathbf{x}_t \in {\mathbb R}^{L\cdot N}$ over time instants $t=0,1,2,...$, drawn i.i.d. in time from a distribution~$ \mathbf{X} = (\mathbf{X}_1,...,\mathbf{X}_N)\sim {\mathbb P}$,   defined over ${\mathbb R}^{L\,N}$. Here, $\mathbf{x}_t = (\mathbf{x}_{1,t},...,\mathbf{x}_{N,t})$, where~$\mathbf{x}_{s,t} \in {\mathbb R}^L$ is a sample of $\mathbf{X}_s$ generated by the MTC device~$s$. The goal is to learn the parameter vector $\mathbf{w}^\star \in {\mathbb R}^Q$ that minimizes the population risk:
  \begin{equation}
  \label{eqn-app-layer-loss}
  \mathbb E_X \left[ \phi(\mathbf w; \mathbf X) \right],
  \end{equation}
where expectation is well-defined for each $\mathbf w \in {\mathbb R}^Q$, and, for each $\mathbf x \in {\mathbb R}^{L\cdot N}$, function $\phi(\cdot;\mathbf x):\,{\mathbb R}^Q \rightarrow \mathbb R$ is differentiable. In the rest of this subsection, we specialize the approach on a single but illustrative example of Consensus; more elaborate examples such as logistic regression are relevant but not included here for brevity. 

\textbf{Example: Consensus -- computing the global average; e.g., \cite{Consensus,DeGroot1974,Kar2009}}. When $w \in {\mathbb R}$, $\mathbf x = (x_1,...,x_N) \in {\mathbb R}^N$ (each MTC device generates scalar data), and $\phi(w,\mathbf x) = \frac{1}{N}\sum_{s=1}^N (x_s-w)^2$, then solving \eqref{eqn-app-layer-loss} corresponds to finding $\frac{1}{N}\sum_{s=1}^N \mathbb E[X_s]$. E.g., when MTC devices are    pollution sensors at different locations in a city, this corresponds to finding the city-wide average pollution.

\textbf{Conventional 3GPP MTC solution}. Consider first the conventional 3GPP MTC system, where samples $\mathbf{x}_t$, $t=0,1,2,...$, arrive (through the communication layer) to a processing unit at the data center at time instants $t=0,1,2,...$ \emph{in their raw form}. (We ignore here the communication delays.) Upon reception of each new sample $\mathbf{x}_t$, the processing unit (destination node~$d$) performs a stochastic gradient update to improve its estimate $\mathbf{w}^{(t)} \in {\mathbb R}^Q$ of $w^\star$:
    \begin{equation}
    \label{eqn-app-layer-stoch-grad}
    \mathbf{w}^{(t+1)} = \mathbf{w}^{(t)} - \eta_t \, \nabla \phi\left( \mathbf{w}^{(t)};\,\mathbf{x}_t\right),
    \end{equation}
where $\nabla \phi\left( \mathbf{w}^{(t)};\,\mathbf{x}_t\right)$ is the gradient of $\phi\left( \cdot ;\,\mathbf{x}_t\right)$ at~$\mathbf{w}^{(t)}$, and $\eta_t$ is the step-size (learning rate). For the consensus example with learning rate $\eta_t = 1/(t+1)$, it can be shown that update~\eqref{eqn-app-layer-stoch-grad} takes the particularly simple form:
\begin{equation}
    \label{eqn-app-layer-stoch-grad-cons}
    {w}^{(t+1)} = \frac{t}{t+1} {w}^{(t)} + \frac{1}{t+1}\left(\frac{1}{N}\sum_{s=1}^N x_{s,t}\right).
    \end{equation}
Note that, with the conventional 3GPP MTC architecture, data samples $\mathbf{x}_t$ are transmitted to the destination node~$d$ for processing in their entirety, i.e., the communication infrastructure acts only as a routing network (forwarder) of the data.

\textbf{Condense solution}. In contrast with the conventional solution, with the Condense architecture the raw data sample $\mathbf{x}_t$ is not transmitted to the data center and is hence not available at the corresponding processing unit. Instead, update~\eqref{eqn-app-layer-stoch-grad} is implemented as follows. Given the current estimate~$\mathbf{w}^{(t)}$, the processing unit (destination node~$d$)
defines function~$f_t(\cdot):=\nabla \phi\left(\mathbf{w}^{(t)};\cdot\right)$. Subsequently, it sends the request to the function processor to perform the decomposition of $f_t(\cdot)$ over the NFC layer. The function processor performs the required decomposition of~$f_t(\cdot)$ into atomic functions and remotely installs the corresponding obtained atomic function at each atomic node (eNB, HeNB, etc.) of the topology.\footnote{We assume that performing decomposition of~$f_t(\cdot)$ and the installation of the atomic functions across the AFC layer is completed prior to the initiation of flow of sample $\mathbf{x}_t$ ``onwards'' through the NFC topology. In other words, the time required for the
latter process is sufficiently smaller than the time intervals of generation of data samples~$\mathbf{x}_t$.} Once the required atomic functions are ready, the sample $\mathbf{x}_t$ starts travelling up the graph~$\mathcal G$, and upon the completion of evaluation of all intermediate atomic functions, the value $f_t(\mathbf{x}_t)$ becomes available at the data center. This in turn means that the processing unit can finalize update~\eqref{eqn-app-layer-stoch-grad}. Specifically, with the consensus example in~\eqref{eqn-app-layer-stoch-grad-cons}, function $f_t(\cdot)$ takes a particularly simple form of the average: $f_t(x)= f(x) = \frac{1}{N} \sum_{s=1}^N x_s$, and it is independent of~$ {w}^{(t)}$ and of~$t$. There exist many simple and efficient methods to decompose\footnote{Strictly speaking, $f_t(x) = \frac{1}{N} \sum_{s=1}^N x_s$ is not defined for the consensus example here as the gradient of $\phi(w,\cdot)$ at $x$, but it is defined as the additive term in (6) which is dependent upon $\mathbf{x}_t$. The gradient actually equals $\frac{1}{N}\sum_{s=1}^N (w - x_s)$;  applying (5) to this gradient form with $\eta_t=1/(t+1)$ yields (6).} the computation of the average, e.g.,~\cite{Aggregation}, and hence algorithm~\eqref{eqn-app-layer-stoch-grad-cons} can be implemented very efficiently within the Condense architecture.

\textbf{Challenges, insights and research directions}. We close this subsection by discussing several challenges which arise when embedding learning algorithms in the Condense architecture. Such challenges are manyfold but are nonetheless already a reality in machine learning practice. First, data arrives in an asynchronous, delayed, and irregular fashion, and it is often noisy. Condense actually embraces this reality and puts the learning task at the center stage: the desired function of the data is of interest, not the data itself. Secondly, it is often the case that, depending on the infrastructure, interface and functionality constraints of the network computation layer, approximations of the desired function computations (as opposed to exact computations) will need to be employed. For instance, function~$f_t(\cdot):=\nabla \phi\left(\mathbf{w}^{(t)};\cdot\right)$ in the example above may only be computable approximately in general. The quality of such an approximation leads to trading-off statistical efficiency of the learning procedure with the accuracy of the network function computation, and the analyses of such trade-offs will be an important research topic. Finally, from a more practical perspective, an important issue is to ensure interoperability with the existing distributed processing paradigms (e.g., Graphlab \cite{Graphlab} and Hadoop \cite{Hadoop}).

\subsection{Neural networks}
With modern large scale applications of neural networks, the number of parameters to be learned (neuron's weights) can be excessively large, like, e.g., with deep neural networks \cite{Bengio2015}. As such, storage of the parameters themselves should be distributed, and their updates also include a large communication cost that needs to be managed \cite{Li2014}.  However, neural networks can be naturally embedded into the Condense architecture -- somewhat similarly to the related work on distributed training for deep learning \cite{Dean2012} -- as detailed next.

Specifically, consider the example of binary classification of the MTC devices' generated data. At each time instant~$t$, $N$ MTC devices generate data vector~$\mathbf{x}_t  = \left( \mathbf{x}_{1,t},...,\mathbf{x}_{N,t} \right) \in {\mathbb R}^{N\cdot L}$, where each device generates an $L$-dimensional vector~$\mathbf{x}_{i,t}$. Each data vector $\mathbf{x}_t$ is associated with its class label~$y_t \in \{-1,1\}$. A binary classifier $F:\,{\mathbb R}^{N\cdot L} \rightarrow \{-1,1\}$ takes a data sample $\mathbf{x}_t$ and generates an estimate $F(\mathbf{x}_t)$ of its class label~$y_t$. Classifier~$F$ is ``learned'' from the available training data $\left( \mathbf{x}_t,\mathbf{y}_t\right)$, $t =0,1,...,T$, where $T$ is the \emph{learning period}. In other words, once the learning period is completed and~$F$ is learned, then the \emph{prediction period} is initiated, and for each new data sample $\mathbf{x}_t$, $t>T$, classifier~$F$ generates an estimate of~$y_t$. 
For example, $\mathbf{x}_t$ can correspond to measurements of pressure, temperature, vibration, acoustic, and other sensors in a large industrial plant within a time period~$t$; $y_t=1$ can correspond to the ``nominal'' plant operation, while $y_t=-1$ to the ``non-nominal'' operation, defined for example as the operation where energy efficiency or greenness standards are not fully satisfied.
     %
     %

We consider neural network-based classifiers~$F$ embedded in the Condense architecture. Therein, the classifier function~$F$ is a composition of the neuron functions associated with each AFC module (node). We consider a Condense rooted tree graph $\mathcal G$ with $N$ sources and one destination node~$d$, however, here we assume $\mathcal G$ is \emph{undirected} as we will need to pass messages upwards and downwards. For convenience, as it is common with neural networks, we organize all nodes in $\mathcal G$ (source, atomic, and the destination node) in levels $\ell = 1,2,...,\mathcal L$, such that the leaves of nodes at the first level ($\ell=1$) are the MTC devices (sources in $\mathcal S$), while the data center's processing unit (the destination node~$d$) corresponds to $\ell = \mathcal L$. Then, all nodes (sources, atomic nodes, and the destination node) are indexed through the index pair~$(\ell,m)$, where $\ell=0,1,...,\mathcal L$ is the level number and $m$ is the order number of a node within its own level, $m=1,...,\mathcal{N}_{\ell}$. Here, $\mathcal{N}_{\ell}$ denotes the number of nodes at the $\ell$-th level.

\textbf{Prediction}. We first consider the \emph{prediction period} $t >T$, assuming that the learning period is completed. This corresponds to actually executing the application task of classification, through evaluating the network function~$F$ at a data sample $\mathbf{x}_t$. (As we will see ahead, the learning period corresponds to \emph{learning} function~$F$, which essentially parallels the task of how a desired network function is decomposed across the AFC modules into the appropriate atomic functions.) Each node~$(\ell,m)$ is assigned a weight vector $\mathbf{w}^{(\ell,m)}$ (obtained within the learning period), whose length equals the number of its associated leaf nodes. Denote by ${x}^{(\ell,m)}_t$ the output (also referred to as activity) of node~$(\ell,m)$ associated with the data sample~$\mathbf{x}_t$, $t > T$, to be computed based on the incoming activities $\mathbf{x}^{(\ell-1,q)}_t$ from the adjacent lower level nodes~$(\ell-1,q)$. Also, denote by $\mathbf{x}^{(\ell-1)}_t$ the vector that stacks all the $\mathbf{x}^{(\ell-1,q)}_t$'s at the level $\ell-1$. Then, ${x}^{(\ell,m)}_t$ is calculated by:
       \begin{equation}
       \label{eqn-NN-1}
       {x}^{(\ell,m)}_t = \mathcal{U} \left( (\mathbf{w}^{(\ell,m)})^\top \mathbf{x}^{(\ell-1)}_t \right),
       \end{equation}
where $z \in \mathbb R \mapsto \mathcal{U}(z) = \frac{1}{1+\mathrm{exp}(z)}$ is the logistic unit function. Therefore, with neural networks, the atomic function~$g^{(\ell,m)}(\cdot)$ associated with each AFC module (node)~$(\ell,m)$ is a composition of 1) the linear map parameterized with its weight vector~$\mathbf{w}^{(\ell,m)}$; and 2) the logistic unit function.

\textbf{Learning}. The learning period corresponds to learning function~$F$, i.e., learning the weight vectors $\mathbf{w}^{(\ell,m)}$ of each AFC module. Differently from the example of minimizing a generic population risk in Subsection~{VI-B}, here learning~$F$ (learning atomic functions of ATC modules) can be done in a distributed way, without the involvement of the FP module. The learning is distributed in the sense that it involves passing messages in the ``upward'' direction (from the MTC devices towards the data center) and the ``downward'' direction (from the data center towards the MTC devices) along the Condense architecture (graph~$\mathcal G$).

Specifically, we assume that weight vectors  $\mathbf{w}^{(\ell,m)}$ are learned by minimizing the log-loss~$J \left( \{ \mathbf{w}^{(\ell,m)}\} \right)$ via a stochastic gradient descent (back-propagation) algorithm, wherein one upward/downward pass corresponds to a single training data sample $\mathbf{x}_t$, $t\leq T$. We now proceed with detailing both the upward and the downward pass \cite{Ripley1996}. We assume that, before initiating the pass, $\mathbf{x}_t$ is available at the bottom-most layer (MTC devices), while label~$y_t$ is available at the data center (destination node~$d$). 
This is reasonable to assume as the label's data size per~$t$ is insignificant (here it is just one bit) and can be delivered to the data center, e.g., by forwarding (conventional) means through the 3GPP MTC system. 

\textbf{Upward pass}. Each node $\left(\ell,m\right)$ computes the gradient of its activity
with respect to its weights as well as with respect to the incoming activities:
\[
\frac{\partial {x}_t^{(\ell,m)}} {\partial \mathbf{w}^{(\ell,m)}}= {x}_t^{(\ell,m)} \left( 1-{ x}_t^{(\ell,m)} \right) \mathbf{x}_t^{(\ell-1)}.
\]

\[
\frac{\partial { x}_t^{(\ell,m)} }{\partial \mathbf{x}_t^{(\ell-1)} }={ x}_t^{(\ell,m)} \left(1-{ x}_t^{(\ell,m)} \right)  \mathbf{w}^{(\ell,m)}.
\]

At this point, node $(\ell,m)$ stores tuple $\left(t,\frac{\partial {x}_t^{(\ell,m)}} {\partial \mathbf{w}^{(\ell,m)}},\frac{\partial {x}_t^{(\ell,m)} }{\partial \mathbf{x}_t^{(\ell-1)} }\right)$
(these are local gradients, needed for weight update in the downward pass.).

\textbf{Downward pass}. Label $y_{t} $ has been received at the data center's processing node. Now gradients of loss function $J$ are backpropagated. Having obtained $\frac{\partial J}{\partial {x}_t^{(\ell,m)}}$, each node $(\ell,m)$ sends to its lower layer neighbour $(\ell-1,k)$ the message consisting of $\left(t,\delta_{t}^{\ell}(m\to k)\right)$ (which we refer to here as gradient contribution), where
\begin{eqnarray*}
\delta_{t}^{\ell}(m\to k) &=& \frac{\partial J}{\partial {x}_t^{(\ell,m)}} \frac{\partial {x}_{t}^{(\ell,m)}}
{\partial \mathbf{x}_{t}^{(\ell-1,k)}} \\
&=& \frac{\partial J} {\partial {x}_{t}^{(\ell,m)}}  {x}_{t}^{(\ell,m)}
\left(1-{x}_{t}^{(\ell,m)} \right) \mathbf{x}_{t}^{(\ell-1)}.
\end{eqnarray*}
Node $(\ell-1,k)$ now can compute
\[
\frac{\partial J}{\partial {x}_{t}^{(\ell-1,k)}}=\sum_{m}\delta_{t}^{\ell}(m\to k).
\]
This is instantiated at the top layer:
\[
\frac{\partial J}{\partial {x}_{t}^{(\mathcal L)}}=-\frac{y_{t}}{ {x}_{t}^{(\mathcal L)}}+\frac{1-y_{t}}{1- {x}_{t}^{(\mathcal L)}}.
\]
Moreover, after sending $\delta_{t}^{\ell}(m\to k)$, node $(\ell,m)$ updates its weights with stochastic gradient update and step-size $\eta_t$:
\begin{eqnarray*}
\mathbf{w}^{(\ell,m)}  \leftarrow   \mathbf{w}^{(\ell,m)}-\eta_t\,
\frac{\partial J}{\partial {x}_{t}^{(\ell,m)}} {x}_{t}^{(\ell,m)} \left(1-{x}_{t}^{(\ell,m)}\right) \mathbf{w}^{(\ell,m)},
\end{eqnarray*}
and removes ``local gradients'' from the memory.

\textbf{Challenges, insights and research directions}. We close this subsection with several challenges and practical considerations which arise when embedding neural networks into the Condense architecture.

The first challenge is on implementing the required AFC modules (atomic functions) in the analog domain.
These modules are typically linear combinations plus nonlinearities (sigmoids, rectified linear units).  Secondly, even when they are implemented in the digital domain, an interesting question is to study the effects of propagation of the quantization error across the Condense architecture.

Next, network topology and busy nodes will dictate that not all nodes see the activities corresponding to the $t$-th example. This is just like the dropout method \cite{Srivastava2014} which deliberately ``switches off'' neurons randomly during each learning stage. Dropout is a hugely successful method for learning regularization as it prevents overfitting by weight co-adaptation and demonstrates that the learning process can be inherently robust to the node failures, echoing the overall case against the learning and network layer separation, which presumes all data to be available on request at all times. Moreover, many activities will not be sent to all the nodes in the upper layer. In this case, $\frac{\partial {x}_{t}^{(\ell,m)}}{\partial {w}^{(\ell, m)}[k]}=0$, so weights will not be affected. In this case, corresponding gradients do not need to be stored, nor does the downward pass need to happen.
More problematic is the situation in which upward pass has happened but downward pass fails at some point, i.e., some of the $\delta_{t}^{\ell}(m\to k)$ are not received at $(\ell-1,k)$. This injects additional noise to the gradient. Studying the effect of this noise is an interesting research topic.

We finally provide some insights on the communication and computational costs. Each AFC module (node) broadcasts one real number per training data example: its activity (together with data example index $t$), in the upward pass, and one real number per example, per receiver: gradient contribution $\delta_{t}^{\ell}(m\to k)$ (together with index $t$), in the downward pass. Thus, each node broadcasts to upper layers and sends specific messages to specific nodes in bottom layers. Upward pass happens whenever a new input is obtained, while downward pass whenever a new output is obtained. Due to this asynchrony, gradient updates might be out of date -- therefore, each node could purge local gradients for outdated examples.


\section{Other implementation aspects}

This Section briefly discusses some aspects of the Condense architecture not considered in earlier Sections.

\textbf{Size of NFC graph~$\mathcal G$}. We first discuss a typical size of an NFC graph. Referring to Figure~6 and assuming a directed rooted tree graph, it can typically have depth (number of layers) around~$5-7$. Regarding the number of source nodes (MTC devices -- lower most layer), it is estimated that the number of MTC devices per macro-cell eNB will be in the range of $10^3-10^5$. The number of small cells per macro cell is in the range of $10^1-10^2$, which makes the number of MTC devices per small cell approximately $10^2-10^3$. Assuming a $30\mathrm{km} \times 30\mathrm{km}$ city area and a $100\mathrm{m} \times 100\mathrm{m}$ coverage of a small cell, we can have a total of $10^4-10^5$ small cells within a city-wide Condense network. Therefore, in a city-wide Condense network, we may have $10^7-10^8$ MTC devices (number of nodes at the lower-most layer), and on the order of $10^4-10^5$ nodes at the (base station) layer above. The number of nodes at the upper layers going further upwards is lower and is few tens or less. In summary, a typical city-wide Condense rooted tree network may have a total of $10^7-10^8$ nodes, it has a large ``width'' and a moderate ``depth''. This goes relatively well in line with the supporting theory; e.g., neural networks are considered deep with depths of order $7$ or so, while arbitrary functions can be well-approximated even with shallow neural networks. Of course, the graph size can be virtually adjusted according to current application needs both horizontally (to adjust width) and vertically (to adjust depth) through implementing multiple (virtual) nodes within a single physical device.

\textbf{Synchronization}. We initially assess that synchronization may not be a major issue with realizing Condense. This is because, actually, synchronization is critical only with implementing analog atomic functions, e.g., within a single HeNB module. Network-wide orchestration of atomic functions may be successfully achieved through the control mechanisms of SDN and NFV, as well as through the usage of \emph{buffering} at the upper Condense layers (HeNB-GW, S-GW, and P-GW), to compensate for delays and asynchrony.

{\renewcommand{\arraystretch}{1.2}
\renewcommand{\tabcolsep}{0.1cm}
\begin{table*}
\centering \caption{Summary of CONDENSE architecture.}
\label{Table_1}
\begin{tabular}{|l|c|l|l|}
\hline
\textbf{Features and pros} & \multicolumn{2}{|c|}{\textbf{Theory and implementation}} & \textbf{Future directions}\\
\hline
- Reconfigurable architecture; & & - Analog: wireless and optical domains; & - Implementation 
   challenges: channel\\
- Novel service of computing & & - Digital: FPGA/software; & estimation, timing and frequency \\
functions over MTC-data; & \textbf{AFC layer} & - Theory: Nomographic functions & offsets and quantization issues;\\
- Three layers: atomic,  & & (analog  wireless) and Reservoir & - Development of standardized A-AFC\\
network and application; & & computing (analog optical) & and D-AFC modules;\\  \cline{2-4}
- Two control elements: topology & & - Topology processor: NFV orchestrator; & - Actual constructions \\
and function processor; & & - Function processor: SDN application; & of function 
  decompositions; \\
- Can be integrated within & \textbf{NFC layer} & - Theory: Network coding for computing, & - Decomposability (solvability) under restricted\\
3GPP MTC architecture; & & sensor fusion and neural networks &  function classes and network topologies; \\
- Can exploit theories of & & & - Development of function and topology \\ 
sensor fusion, network coding and & & & processor SDN/NFV modules;\\ \cline{2-4}
computation and neural networks; & & - Implementation examples: RLNC, neural & - Asynchronous, delayed and irregular\\
- Can be customized for variety & \textbf{Application} & networks and stochastic gradient descent; & arrival of data;\\ 
 of MTC applications;& \textbf{layer} & - Theory: neural networks, statistical & - Inexact network function computation;\\
 & & learning and prediction & - Interoperability with existing data\\
 & & & analytics platforms; \\
 \hline 
\end{tabular}
\end{table*}}

\textbf{Communication, computational, and storage costs}. We now discuss reductions of communication costs (per application task) of Condense with respect to the conventional (forwarding) 3GPP MTC solution. How much communications is saved depends largely on the application at hand. For very simple tasks (functions), like, e.g., computing maximum or global average, it is easy to see that the savings can be very high. In contrast, for forwarding (computing the identity function), the savings may not be achieved (but the reliability is improved through random linear network coding). Also, overall communication savings depend on the overhead incurred by the signalling from the topology and function
processors to the AFC modules (in order to orchestrate the topology, perform function decomposition and instal the appropriate atomic functions, etc.) However, this overhead is projected to eventually become small, as, upon a significant development of  the technology, atomic function libraries and NFC
decompositions will be pre-installed. Further, it is clear that Condense requires additional storage
and computational functionalities at network nodes, when compared with current 3GPP MTC systems. However, this is a reasonable assumption for the modules (eNBs, GWs, etc.) of 3GPP MTC systems due to upcoming trends in mobile edge computing \cite{MEC}.

\textbf{Data privacy and data loss}.
Condense naturally improves upon privacy of IoT systems, as the data center (except when computing the identity function) does not receive the data in its raw from. Finally, if certain IoT-generated data has to be stored in a cloud data center in its raw form so as to ensure its long lifetime, Condense supports this functionality through identity functions. However, it is natural to expect that this request is, on average across all IoT data sources and all applications, only occasionally imposed, rendering significant communication savings overall.

Finally, Table \ref{Table_1} provides a summary of the proposed architecture. The table briefly indicates main points presented in this paper in terms of the advantages of the proposed architecture, relevant theoretical and implementation aspects, and main future research directions. 

\section{Conclusions}

In this paper, we proposed a novel architecture for knowledge acquisition of IoT-generated data within the 3GPP MTC (machine type communications) systems, which we refer to as Condense. The Condense architecture introduces a novel service within 3GPP MTC systems -- computing linear and non-linear functions over the data generated by MTC devices. This service brings about the possibility that the underlying communication infrastructure communicates only the desired function of the MTC-generated data (as required by the given application at hand), and not the raw data in its entirety. This transformational approach has the potential to dramatically reduce the pressure on the 3GPP MTC communication infrastructure. 

The paper provides contributions along two main directions. First, from the architectural side, we describe in detail how the function computation service can be realized within 3GPP MTC systems. Second, from the theoretical side, we survey the relevant literature on the possibilities of realizing ``atomic'' functions in both analog and digital domains, as well as on the theories and techniques for function decomposition over networks, including the literature on sensor fusion, network coding for computing, and neural networks. The paper discusses challenges, provides insights, and identifies future research directions for implementing function computation and function decomposition within practical 3GPP MTC systems.

\section*{Acknowledgment}
The authors thank the following researchers for valuable help in developing the Condense concept: J. Coon, R. Vicente, M. Greferath, O. Gnilke, R. Freij-Hollanti, A. Vazquez Castro, V. Crnojevic, G. Chatzikostas, C. Mirasso, P. Colet, and M.~C. Soriano. The authors would also like to acknowledge valuable support for collaboration through the EU COST IC 1104 Action.

\end{document}